\documentclass[a4paper,fleqn]{cas-sc}

\hypersetup{
	colorlinks=true,
	allcolors=blue
}

\usepackage[numbers,sort&compress]{natbib}

\def\tsc#1{\csdef{#1}{\textsc{\lowercase{#1}}\xspace}}
\tsc{WGM}
\tsc{QE}

\usepackage{amsmath}
\usepackage[T1]{fontenc}

\usepackage{setspace}

\usepackage{pdflscape}
\usepackage{float}
\usepackage{lscape}
\usepackage{multirow}
\usepackage{booktabs}
\usepackage{graphicx}
\usepackage{subcaption}
\usepackage{rotating}
\usepackage{array}
\usepackage{longtable}
\usepackage{makecell} 
\usepackage{doi}

\usepackage{anyfontsize}

\usepackage{etoolbox}
\AtBeginEnvironment{table}{\fontsize{10pt}{12pt}\selectfont}
\AtBeginEnvironment{longtable}{\fontsize{10pt}{12pt}\selectfont}

\ExplSyntaxOn
\AtBeginDocument{
	\cs_gset:Npn \__first_footerline:
	{ \group_begin: \small \sffamily \__short_authors: \hfill arXiv~preprint \group_end: }
}
\ExplSyntaxOff

\begin{document}
\let\WriteBookmarks\relax
\def\floatpagepagefraction{1}
\def\textpagefraction{.001}

\shorttitle{SSL and ViViT for Audio MOS Prediction}    

\shortauthors{M.O. Duman et al.}  

\title [mode = title]{WTMOS: A Robust WavLM-Transformer Framework and ViViT Benchmarking for Cross-Corpus Audio MOS Prediction} 



%

\author[1]{Mustafa Ozan Duman}[orcid=0009-0001-7017-8758]

\cormark[1]


\ead{duman@utia.cas.cz}


\credit{Conceptualization, Software, Data Curation, Writing - Original Draft}

\affiliation[1]{organization={The Czech Academy of Sciences},
	department={Institute of Information Theory and Automation},
	city={Prague},
	postcode={18200},
	country={Czech Republic}}

\affiliation[2]{organization={Bursa Uludag University},
	department={Computer Engineering},
	city={Bursa},
	postcode={16240},
	country={Turkiye}}

\author[2]{Ahmet Emir Dirik}[orcid=0000-0002-6200-1717]


\ead{edirik@uludag.edu.tr}


\credit{Supervision, Methodology, Writing - Review \& Editing}


\cortext[1]{Corresponding author}



\begin{abstract}
Automatic Mean Opinion Score (MOS) prediction is essential for evaluating large-scale synthetic speech and audio enhancement systems, yet models frequently struggle with domain shift. This study presents WTMOS, a robust WavLM-Transformer framework and Video Vision Transformer (ViViT) benchmarking suite across three architectural configurations: Frozen Self-Supervised Learning (SSL-FRZ), Fine-Tuned SSL (SSL-FT), and ViViT. Evaluation is conducted in two phases: Part I utilizes a consolidated corpus of 130,000 samples across 19 diverse datasets, while Part II focuses on a purified 17-dataset English-only corpus. To assess robustness, a systematic Leave-One-Dataset-Out (LODO) protocol is employed to quantify the generalization gap between seen and unseen distributions. Finally, the top-performing model is benchmarked against 18 state-of-the-art (SOTA) metrics using the ARECHO framework.

Results demonstrate that an English-only purified corpus consistently yields higher predictive precision across all architectures. While SSL-FT achieves the highest performance on seen validation data, the SSL-FRZ model provides superior robustness on unseen distributions, achieving a competitive Mean Squared Error (MSE) of 0.36 on the URGENT 2024 benchmark—closely matching domain-optimized SOTA metrics (MSE 0.30). Although the ViViT architecture remains below SSL-based models in total capacity, it delivers stable results in English-only trials. LODO results confirm that while models perform significantly better on seen samples, frozen SSL embeddings combined with deep Transformer encoders offer the most stable and scalable solution for universal speech quality assessment. To support further research, the top-performing English-only SSL-Transformer model and weights are made publicly available via Hugging Face.
\end{abstract}




\maketitle

\section{Introduction}\label{sec:introduction}
MOS serves as the definitive human-centric metric for evaluating the perceptual quality of multimedia content. Traditionally, MOS is a subjective measure where human listeners rate audio samples on a discrete scale, typically from 1 (Poor) to 5 (Excellent). In the current era of large-scale data driven by advancements in speech synthesis, voice conversion, and deepfake detection, the quality of acoustic data is paramount \cite{ITU_P800}.

However, as datasets scale into the hundreds of thousands of samples, manual subjective scoring becomes logistically and financially infeasible. This necessitates the development of computational prediction models capable of automatically and accurately estimating MOS \cite{NEED_MOS_PREDICTER}. Literature categorizes these advancements into traditional machine learning approaches, deep neural network (DNN) based methods, and more recently, self-supervised learning (SSL) and unsupervised frameworks \cite{MOS_PREDICTER_TYPES}.

In MOS prediction, Mel-Frequency Cepstral Coefficients (MFCCs) are widely used to represent the short-term power spectrum of sound, capturing spectral characteristics that align closely with human auditory perception. These coefficients focus on the timbre and spectral envelope of the audio, providing valuable features for predicting speech quality \cite{MFCC_EXPLANATION}. However, the extraction process involves segmenting audio into small, consecutive windows and applying a Fast Fourier Transform (FFT) to each. This discretization, while computationally efficient, inevitably leads to information loss; phase information is typically discarded, and the temporal resolution is strictly limited by the chosen window size. Furthermore, MFCCs suffer from variability in matrix size depending on audio duration, presenting a challenge for architectures like CNNs or vision transformer (ViT)-based models which typically require fixed-size inputs \cite{MOD_CONFERENCE}.

To address these constraints while preserving the human-centric benefits of spectral data, we employ ViViT \cite{VIVIT_EXPLANATION}-Transformer Encoder architecture. In this approach, MFCCs are extracted from 5-second segments, using zero-padding and masking to ensure structural consistency across varying audio lengths. By utilizing a classification (CLS) token within the Transformer Encoder to aggregate information across these segments, the model effectively captures global quality features from the 2D spectral-temporal representation.

As a prominent alternative to spectral features, SSL representations have gained traction for their ability to process raw audio data directly via 1D convolutional filters \cite{WAVLM}. Unlike MFCCs, which rely on a human-centric mathematical prior, SSL models learn hierarchical features that capture deep contextual and linguistic nuances. To fully exploit these learned embeddings, we integrate the SSL backbone with a downstream Transformer Encoder. This configuration allows the model to apply self-attention mechanisms across the high-dimensional feature sequences extracted by the SSL model, effectively modeling long-range dependencies and temporal nuances that are often lost in traditional pooling layers. While pairing SSL with simple linear heads is common in literature, the use of a deep Transformer Encoder to refine SSL-based MOS prediction remains under-explored, especially when evaluating the trade-offs between frozen versus fine-tuned backbones at a massive scale. Determining whether the significant overhead of fine-tuning these complex architectures is justified for 100,000+ utterances remains a critical question for developing robust, real-world MOS predictors.

\begin{figure} 
	\centering
	\includegraphics[width=\textwidth]{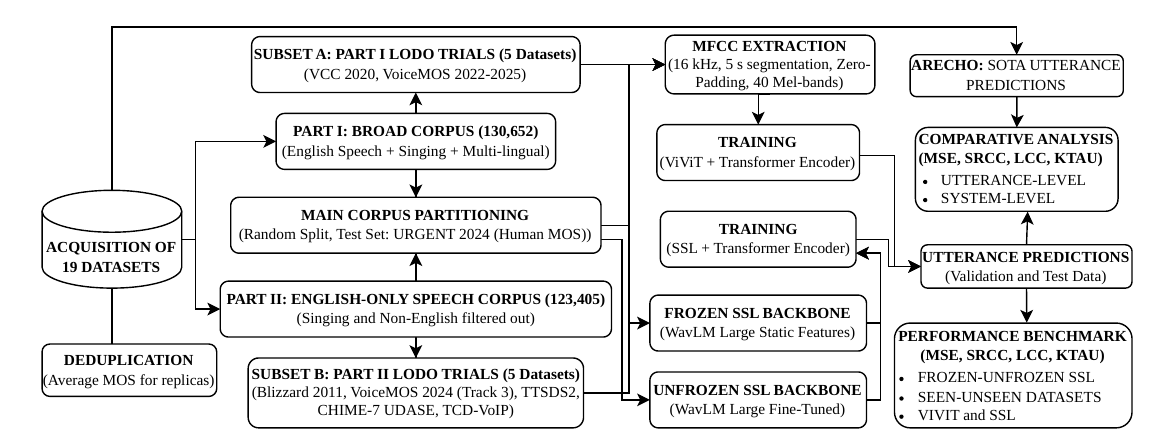}
	\caption{Proposed experimental pipeline for large-scale MOS prediction and multi-metric benchmarking. The framework illustrates the flow from data acquisition and deduplication to parallel model training (SSL-Transformer and ViViT) and final evaluation against ARECHO SOTA.}
	\label{fig:overall_system}
\end{figure}

While previous research has achieved significant benchmarks using MFCC and/or SSL representations, several critical gaps remain. Most existing models are often optimized for narrow domains and struggle with generalization across diverse, unseen datasets \cite{NISQA, UTMOS, SHEET}. Specifically, there is a lack of comparative research evaluating the efficiency of ViViT-Transformer architectures against SSL-Transformer configurations at the scale of 130,000+ utterances. While SSL models are often used as static feature extractors, the potential of a unified SSL-Transformer Encoder framework—and how it behaves when the SSL backbone is fine-tuned versus frozen—has not been fully documented in the context of cross-domain, multilingual, and singing audio \cite{PESQ, VISQOLv3, PYSEPM, DNS_P808,DNS_P835, NISQA, PLCMOS, UTMOS, UTMOSv2, MOS-BENCH, SHEET, SINGMOS,  MetaAudioboxAesthetics,  NORESQA, NORESQA-MOS,   SCOREQ, NOMAD}.

This study addresses the aforementioned gaps through an unprecedented evaluation scale and a rigorous two-part experimental framework designed to contrast human-centric spectral priors with high-capacity learned embeddings. The comprehensive workflow of this research is illustrated in Figure \ref{fig:overall_system}.

In Part I, we utilize an extensive corpus of 130,652 utterances spanning 19 datasets to conduct a head-to-head evaluation between a novel ViViT-Transformer architecture and an SSL-Transformer Encoder. Within the SSL-based path, we investigate two distinct training regimes: an unfrozen (fine-tuned) backbone to capture domain-specific acoustic nuances and a frozen backbone to assess the generalizability of static self-supervised features. To quantify the robustness of these models against unseen data, we implement a systematic 6-case controlled experiment following LODO protocol. In this setup, we isolate five representative datasets to serve as independent test targets across five cases, while the sixth case utilizes a randomized seen split to establish a performance baseline. This allows for a precise measurement of the performance ceiling within familiar distributions compared to the degradation encountered in out-of-distribution scenarios.

In Part II, recognizing that most SOTA models are biased toward English speech, we isolate a 123,405-sample English-Only corpus. By filtering out approximately 7,000 samples containing singing and multilingual audio from the original collection, we create a linguistically purified subset. By replicating the LODO experimental protocol—including both fine-tuned and frozen SSL configurations—on this subset, we analyze performance shifts and domain-transfer capabilities, specifically identifying how models respond when the acoustic data is restricted to standard English speech versus the diverse acoustic profiles of the broad corpus.

To ensure robust validation, we employ the ARECHO framework \cite{ARECHO} as a standalone SOTA benchmark. Unlike traditional toolkits, ARECHO functions as a reference-free labeler that we leverage to generate specialized quality profiles for our entire dataset. By obtaining predictions across 18 distinct metrics, we facilitate a multi-dimensional comparative analysis. Crucially, our evaluation is conducted at both the utterance level (to measure sample-specific accuracy) and the system level (to assess the model's ability to rank different datasets), benchmarking our predictions against established SOTA metrics using MSE, Linear Correlation Coefficient (LCC), Spearman’s Rank Correlation Coefficient (SRCC), and Kendall’s Tau (KTAU).

The primary contributions of this work are summarized as follows:

\begin{itemize}
	\item \textbf{Extensive Benchmarking:} We provide a large-scale evaluation of MOS prediction robustness leveraging a consolidated corpus of 130,652 files across 19 diverse datasets.
	\item \textbf{Generalization Analysis:} We quantify the performance gap between seen and unseen domains through a systematic LODO evaluation, highlighting the persistent challenges of domain shift in automated quality assessment.
	\item \textbf{Architectural Comparison:} We introduce and contrast three sophisticated frameworks: a ViViT-Transformer for spectral 2D MFCCs, and both Frozen and Fine-Tuned SSL-Transformer Encoders. Our results demonstrate the stability of spectral-priors versus the superior predictive power of learned embeddings.
	\item \textbf{Cross-Domain \& Language Insights:} We document the performance trade-offs between broader multilingual corpora and English-specific specialization, specifically identifying the impact of high-variance audio like singing on prediction accuracy.
	\item \textbf{SOTA Validation:} We achieve results that surpass or match 18 established SOTA models within the ARECHO benchmark, with our best model achieving a competitive 0.36 MSE on the URGENT 2024 test set.
	\item \textbf{Open-Source Contribution:} We release the complete implementation and top-performing model checkpoints to the research community via GitHub and Hugging Face to facilitate reproducible speech quality assessment.\footnote{Code: \url{https://github.com/mustafa-ozan/audio_mos_prediction_SSL_ViViT_codes}; Model: \url{https://huggingface.co/mustafa-ozan-duman/wavlm-transformer-mos-english}}
	
\end{itemize}

The remainder of this paper is organized as follows: Section 2 provides the  Related Work, reviewing the evolution of objective speech quality metrics and the current landscape of SSL-based benchmarks. Section 3 details our Methodology, including the specific architectural implementations of the ViViT-Transformer and SSL-Transformer frameworks, alongside the 6-case experimental protocol. Section 4 presents the results and Section 5 provides a Discussion of the findings regarding generalization, domain shifts, and the trade-offs of fine-tuning. Section 6 concludes the study with the Conclusion and introduces the finalized public model.

\section{Related works}\label{sec:related work}

Historically, speech quality assessment relied on full-reference metrics that required a clean signal for comparison. Standards such as PESQ \cite{PESQ} utilized psychoacoustic auditory transforms to map raw waveforms into loudness representations for telephony networks. More recently, ViSQOL v3 \cite{VISQOLv3} modernized this approach by using gammatone spectrograms and the Neurogram Similarity Index Measure to improve robustness against temporal warping and low-bitrate codecs. Early efforts to automate these judgments, such as the Hu and Loizou study \cite{PYSEPM}, demonstrated that composite measures outperformed individual signal-based assessments on the NOIZEUS corpus.

The shift toward non-intrusive (no-reference) models was spearheaded by CNN-based architectures. Microsoft’s DNSMOS (P.808 and P.835) \cite{DNS_P808, DNS_P835} frameworks introduced large-scale, self-teaching models that process log-power spectrograms to predict both overall quality and multidimensional scores like signal (SIG) and background noise (BAK). Similarly, NISQA \cite{NISQA} advanced this by incorporating Self-Attention mechanisms and Attention-Pooling to capture temporal recency effects and network-induced degradations across 81 datasets. Specialized models like PLCMOS \cite{PLCMOS} have further refined these techniques, using micro-augmentations to specifically target packet-loss concealment artifacts with state-of-the-art accuracy.

The integration of SSL backbones has significantly improved the extraction of high-dimensional acoustic features from raw waveforms. The UTMOS \cite{UTMOS} and UTMOSv2 \cite{UTMOSv2} systems demonstrated the power of ensembling multiple SSL learners, with the latter uniquely incorporating an EfficientNetV2 image classifier to detect subtle spectral artifacts. To address the challenge of generalization, frameworks like MOS-Bench \cite{MOS-BENCH} and the SHEET \cite{SHEET} toolkit provide unified recipes for evaluating models across diverse, out-of-domain datasets. These advancements have enabled models to scale beyond pure speech; for instance, SingMOS \cite{SINGMOS} established a dedicated benchmark for singing voice quality, while Meta Audiobox Aesthetics \cite{MetaAudioboxAesthetics} expanded assessment into the four-axis aesthetic evaluation of music and general sound.

Current research is exploring reference-invariant models that do not require an exact clean counterpart. NORESQA \cite{NORESQA} and NORESQA-MOS \cite{NORESQA-MOS} pioneered the use of non-matching references through multi-task learning, allowing the model to ground MOS estimation by comparing content-dissimilar recordings. Similarly, SCOREQ \cite{SCOREQ} and NOMAD \cite{NOMAD} leverage unsupervised triplet loss functions to disentangle speech content from degradation intensity. These approaches, often evaluated on diverse corpora like TCD-VoIP \cite{TCD-VoIP} and BVCC \cite{BVCC}, offer a scalable path forward for quality assessment in the absence of human-labeled data.

\section{Method}\label{sec:method}

In this section, we detail the multi-stage experimental design of our study. We first describe the acquisition and curation of the 130,000-utterance corpus. Subsequently, we present the architectural specifications of the proposed SSL-Transformer and ViViT-MFCC frameworks. Finally, we outline the systematic experimental protocols used to evaluate model generalization and domain robustness.

\subsection{Dataset acquisition}\label{subsec:dataset}

The finalized corpus for Part I consists of 130,652 unique utterances. For Part II, a refined subset of 123,405 utterances was used after excluding singing and non-English samples. Table \ref{tab:datasets} summarizes the total distribution.

\begin{table}[!htbp]
	\centering
	\footnotesize 
	\caption{Detailed composition of the unified audio MOS dataset including sample counts and domain types. The repeated and empty parts omitted while constructing Part I corpus. VoiceMOS 2023 track 2, VoiceMOS 2024 track 2, and non-english part of VCC2020 omitted while constructing Part II corpus.}\label{tab:datasets}
	
		\begin{tabular}{l l r} 
			\toprule
			\textbf{Dataset Source} & \textbf{Link} & \textbf{Number of Samples} \\
			\midrule
			Blizzard Challenge 2008 \cite{BLIZZARD2008}& \href{https://www.cstr.ed.ac.uk/projects/blizzard/data.html}{Link} & 882 \\
			Blizzard Challenge 2009 \cite{BLIZZARD2009}& \href{https://www.cstr.ed.ac.uk/projects/blizzard/data.html}{Link} & 1,450 \\
			Blizzard Challenge 2010 \cite{BLIZZARD2010}& \href{https://www.cstr.ed.ac.uk/projects/blizzard/data.html}{Link} & 1,552 \\
			Blizzard Challenge 2011 \cite{BLIZZARD2011}& \href{https://www.cstr.ed.ac.uk/projects/blizzard/data.html}{Link} & 507 \\
			Blizzard Challenge 2012 \cite{BLIZZARD2012}& \href{https://www.cstr.ed.ac.uk/projects/blizzard/data.html}{Link} & 442 \\
			Blizzard Challenge 2013 \cite{BLIZZARD2013}& \href{https://www.cstr.ed.ac.uk/projects/blizzard/data.html}{Link} & 1,296 \\
			Blizzard Challenge 2013 Extension \cite{BLIZZARD2013_EXT}& \href{https://www.cstr.ed.ac.uk/projects/blizzard/data.html}{Link} & 353 (210 repeat) \\
			NISQA Corpus (TEST LIVE TALK excluded)\cite{NISQA}& \href{https://github.com/gabrielmittag/NISQA/wiki/NISQA-Corpus}{Link} & 14,200 \\
			PSTN \cite{PSTN}& \href{https://github.com/ConferencingSpeech/ConferencingSpeech2022/tree/main/Training/Dev\%20datasets}{Link} & 58,709 \\
			SOMOS \cite{SOMOS}& \href{https://zenodo.org/records/7378801}{Link} & 20,100 \\
			VCC 2020 \cite{VCC2020_dataset}& \href{https://github.com/nii-yamagishilab/VCC2020-listeningtest}{Link} & 6,090 (1 empty - 18 non-English) \\
			VoiceMOS 2022 \cite{VOICEMOS2022}& \href{https://zenodo.org/records/6572573\#.Yphw5y8RprQ}{Link} & 7,106 \\
			VoiceMOS 2023 (Track 2) \cite{VOICEMOS2023}& \href{https://sites.google.com/view/voicemos-challenge/resources}{Link} & 4,040 \\
			VoiceMOS 2024 (Track 1) \cite{VOICEMOS2024}& \href{https://sites.google.com/view/voicemos-challenge/resources}{Link} & 1,500 (all repeated) \\
			VoiceMOS 2024 (Track 2) \cite{VOICEMOS2024}& \href{https://www.codabench.org/competitions/2650/}{Link} & 3,189 \\		
			VoiceMOS 2024 (Track 3) \cite{VOICEMOS2024}& \href{https://sites.google.com/view/voicemos-challenge/resources}{Link} & 380\\
			VoiceMOS 2025 (Track 2(TTS)-3) \cite{VOICEMOS2025}& \href{https://sites.google.com/view/voicemos-challenge/resources}{Link} & 2,135 \\
			URGENT 2024 \cite{URGENT2024}& \href{https://huggingface.co/datasets/urgent-challenge/urgent2024_mos}{Link} & 6,900 \\
			TTSDS2 \cite{TTSDS2}& \href{https://huggingface.co/datasets/ttsds/listening_test}{Link} & 460 \\
			CHiME-7 \cite{CHIME7_UDASE}& \href{https://zenodo.org/records/10418311}{Link} & 688 \\
			TCD-VoIP \cite{TCD-VoIP}& \href{https://drive.google.com/file/d/1rHJN34vP-W8SJtjpNUnx5RIks3o5L5he/view}{Link} & 384 \\
			\midrule
			\textbf{Total Part I (Broad Corpus)} & & \textbf{130,652} \\
			\textbf{Total Part II (English Only)} & & \textbf{123,405} \\
			\bottomrule
		\end{tabular}
\end{table}

\subsubsection{Blizzard challenge}

We extracted naturalness MOS ratings from the Blizzard Challenge series from 2008 to 2013 \cite{BLIZZARD2008, BLIZZARD2009, BLIZZARD2010, BLIZZARD2011, BLIZZARD2012, BLIZZARD2013}. We strictly filtered for English-language tasks, including Received Pronunciation, Southern British, and Arctic accents. For the Blizzard 2013 extension \cite{BLIZZARD2013_EXT}, we included both original utterances and extension files, applying arithmetic averaging to MOS labels for any repeated waveforms. Regarding other years, we found results for Blizzard 2016 \cite{BLIZZARD2016} were only shared as averages. We communicated with the corresponding organizers to request utterance-level labels, but they stated that these results are no longer accessible given the time passed. Other challenges in the series were either conducted in non-English languages or the results were not publicly shared.

\subsubsection{Voice conversion challenge}

The VCC 2020 dataset \cite{VCC2020, VCC2020_dataset} was utilized for its intra-lingual semiparallel and cross-lingual VC tasks. We extracted utterance-level MOS labels for the English tasks. We excluded 18 non-English (Japanese) target speaker reference files from our final pure English subset. For VCC 2016 and 2018, the available results are only at the system level rather than the utterance level. We attempted to contact the organizers for more granular data, but received no response. Results for the 2023 and 2025 challenges have not yet been publicly shared.

\subsubsection{VoiceMOS challenge}
The VoiceMOS series represents a central pillar of our dataset, as these challenges specifically aim to bridge the gap between objective metrics and human perception in synthesized speech. We carefully filtered these challenges to maintain our linguistic and domain-specific focus:

VoiceMOS 2022: This inaugural challenge was founded to encourage research in automatic MOS prediction for Text-to-Speech (TTS) and VC systems. It featured a Main track and an Out-of-Domain (OOD) track. We strictly utilized the Main track data, which provides a large-scale collection of MOS ratings for a wide variety of synthesis systems spanning many years. The OOD track was excluded from our primary training as it consisted of Blizzard 2019 data, which is conducted in Mandarin and thus falls outside our English-centric scope \cite{BVCC, VOICEMOS2022}.

VoiceMOS 2023: This challenge emphasized zero-shot OOD MOS prediction across three scenarios. We selected Track 2, which focuses on singing voice conversion. This track includes diverse audio with singing, so in part 2 we extracted it from main dataset to get pure english speech dataset \cite{VOICEMOS2023}.

VoiceMOS 2024: This iteration included three distinct tracks, all of which were integrated into our study with specific considerations. Track 1 focused on zoomed-in high-quality systems; however, we identified that much of this data overlaps with the VoiceMOS 2022 set, requiring the rigorous deduplication and label-averaging process described in Section 3.2. Track 2 provided an extension of singing voice synthesis with a larger variety of systems and languages. Given its multi-lingual nature, we filtered it out for Part II. Track 3 involved semi-supervised MOS prediction for noisy, clean, and enhanced speech; we utilized the available MOS-labeled data from this track to bolster our model's robustness against real-world environmental distortions \cite{VOICEMOS2024}.

VoiceMOS 2025: For the most recent challenge, we utilized Track 2 and Track 3. From Track 2, which utilizes the Meta Audiobox Aesthetics framework to assess production quality and enjoyment, we exclusively extracted the TTS subset, as the text-to-audio and text-to-music components were not relevant to our vocal quality task. Track 3 was selected for its unique focus on high sampling frequencies (16kHz, 24kHz, and 48kHz), allowing us to train and evaluate our models on full-band audio. Track 1 (Text-to-Music) was entirely excluded as it pertains to music expert evaluations rather than speech quality assessment \cite{VOICEMOS2025}.

\subsubsection{URGENT challenge}

The URGENT (Universality, Robustness, and Generalizability for EnhancemeNT) series focuses on building universal speech enhancement models \cite{URGENT_CHALLANGE}. These challenges are critical to our study, particularly for evaluating unseen performance. We utilized the evaluation dataset from URGENT 2024 challenge, which consists of English speech samples processed to handle different distortions and sampling frequencies across varied acoustic environments. This dataset includes ground-truth human MOS labels and serves as our primary test set for assessing model generalizability \cite{URGENT2024}.

While data for the URGENT 2025 challenge is available, we excluded it from our training and ground-truth evaluation because its MOS labels are derived from SOTA model predictions rather than real human subjective ratings \cite{URGENT2025}.

URGENT 2026 challenge introduced Track 2, focused on Speech Quality Assessment (SQA). While the 2026 challenge itself does not provide new human MOS labels directly for its primary task, its framework incorporates several significant standalone datasets as TTSDS2, CHiME-7 UDASE, TCD-VoIP, which we have treated as individual components of our corpus \cite{URGENT2026}.

\subsubsection{Other used datasets}

In addition to the challenge-based data, several standalone corpora were integrated to cover specific synthetic speech:

TTSDS2 (Text to Speech Distribution Score 2): An improved resource for evaluating synthetic speech that is close to real speech. We utilized this dataset's 11,000+ subjective ratings collected for 19 synthetic systems. The ratings include MOS, Comparative MOS, and Speaker Similarity MOS. It is particularly valuable for its ability to distinguish between high-quality modern TTS outputs \cite{TTSDS2}.

CHiME-7 UDASE: Originating from the 7th CHiME challenge (Unsupervised Domain Adaptation for Speech Enhancement), this dataset consists of real-world multi-speaker conversational recordings made in noisy and reverberant domestic environments. It provides a challenging test domain where synthetic training data often fails to reflect real-world acoustic complexities \cite{CHIME7, CHIME7_UDASE}.

TCD-VoIP: This dataset targets degradations occurring in Voice over IP calls that are independent of the specific codec or hardware. It contains speech samples with common VoIP artifacts and subjective opinion scores from 24 listeners, serving as a vital resource for benchmarking communication system quality \cite{TCD-VoIP}.

PSTN (The Public Switched Telephone Network): PSTN training dataset provides a significant volume of English-language data, totaling 58,709 clips with a uniform duration of 10 seconds. The corpus is divided into 40,739 files based on noisy reference signals and 17,970 files based on clean references, offering a robust foundation for modeling telephony-specific distortions \cite{PSTN}.

NISQA Corpus: This dataset includes over 14,000 speech samples designed to evaluate distortions in communication networks. It covers both simulated conditions (codecs, packet-loss, background noise) and live conditions (mobile calls and VoIP platforms like Zoom, Skype, and WhatsApp). Each sample is labeled with 97,000 total human ratings covering overall MOS and four specific dimensions: Noisiness, Coloration, Discontinuity, and Loudness. For our study, we excluded the German-language NISQA TEST LIVETALK section to maintain linguistic consistency \cite{NISQA}.

SOMOS: The Samsung Open MOS dataset focus on neural TTS naturalness. It contains 20,000 synthetic utterances generated by Tacotron-like acoustic models and an LPCNet vocoder trained on the LJ Speech dataset. The 2,000 text sentences used for synthesis were selected from diverse sources, including Blizzard Challenge texts (2007–2016) and Wikipedia. Naturalness evaluations were crowdsourced from listeners in the US, GB, and CA locales, providing a highly reliable set of evaluations for high-quality synthetic speech \cite{SOMOS}.

\subsection{Pre-processing and feature extraction}\label{subsec:data preprocessing}

The primary challenge of consolidating 19 different datasets is the variation in native sampling rates, which range from narrowband (8 kHz) to high-fidelity (48 kHz). To ensure architectural compatibility with the WavLM-Large backbone—which was pre-trained on 16 kHz audio—we perform on-the-fly resampling during the data loading stage. Every waveform is loaded directly at a target sampling rate of 16 kHz. For signals with higher native rates, this acts as a high-quality downsampling process that preserves the most perceptually relevant frequencies for MOS prediction while ensuring the input aligns with the SSL model’s convolutional front-end.

Once loaded at 16 kHz, each audio signal is partitioned into fixed-length segments of 5 seconds ($80,000$ samples). Utterances longer than 5 seconds are divided into multiple sequential frames. If the final segment or the entire utterance is shorter than the 5-second threshold, zero-padding is applied to the end of the signal to maintain a consistent matrix shape. This standardization is critical for the ViViT-MFCC model, which treats these segments as video frames of equal dimensions.

The 16 kHz segments are then processed through two parallel feature extraction paths. For the MFCCs used in ViViT architecture, we extract 40 Mel-bands using an FFT size of 512 and a hop length of 256. This creates a 2D spectral-temporal representation of each 5-second window. During this extraction, a segment-level metadata is generated to track the valid frame count, enabling the attention mechanism to ignore the padded regions via a source key padding mask.

On the other hand, the raw 16 kHz standardized waveforms are fed to the WavLM-Large backbone. This generates a sequence of 1024-dimensional latent vectors, where each vector corresponds to approximately 20ms of audio. By loading at 16 kHz, we ensure that the temporal resolution of these embeddings correctly reflects the acoustic characteristics the SSL model was trained to recognize.

\subsection{Model architectures}\label{subsec:models}

\subsubsection{SSL-Transformer encoder architecture}

The effectiveness of a SSL model for MOS prediction depends on its ability to disentangle speech content from environmental and synthetic degradations. While architectures like wav2vec 2.0 \cite{WAV2VEC2} and HuBERT \cite{HUBERT} have set benchmarks in Automatic Speech Recognition, their pre-training objectives are primarily optimized for capturing linguistic and phonetic information. In contrast, we selected WavLM-Large as our backbone due to its unique Masked Speech Denoising and Prediction framework. Unlike its predecessors, WavLM is trained not only to predict masked speech units but also to reconstruct clean latent representations from simulated noisy or overlapping speech. This denoising objective forces the model to learn paralinguistic and acoustic features such as noise, reverberation, and speaker identity that are vital for quality assessment. Furthermore, WavLM-Large \cite{WAVLM} incorporates a gated relative position bias in its Transformer structure, which enhances its sensitivity to the temporal sequence of speech, allowing it to better identify localized artifacts like jitter or packet loss that often degrade Mean Opinion Scores.

\begin{figure} 
	\centering
	\includegraphics[width=\textwidth]{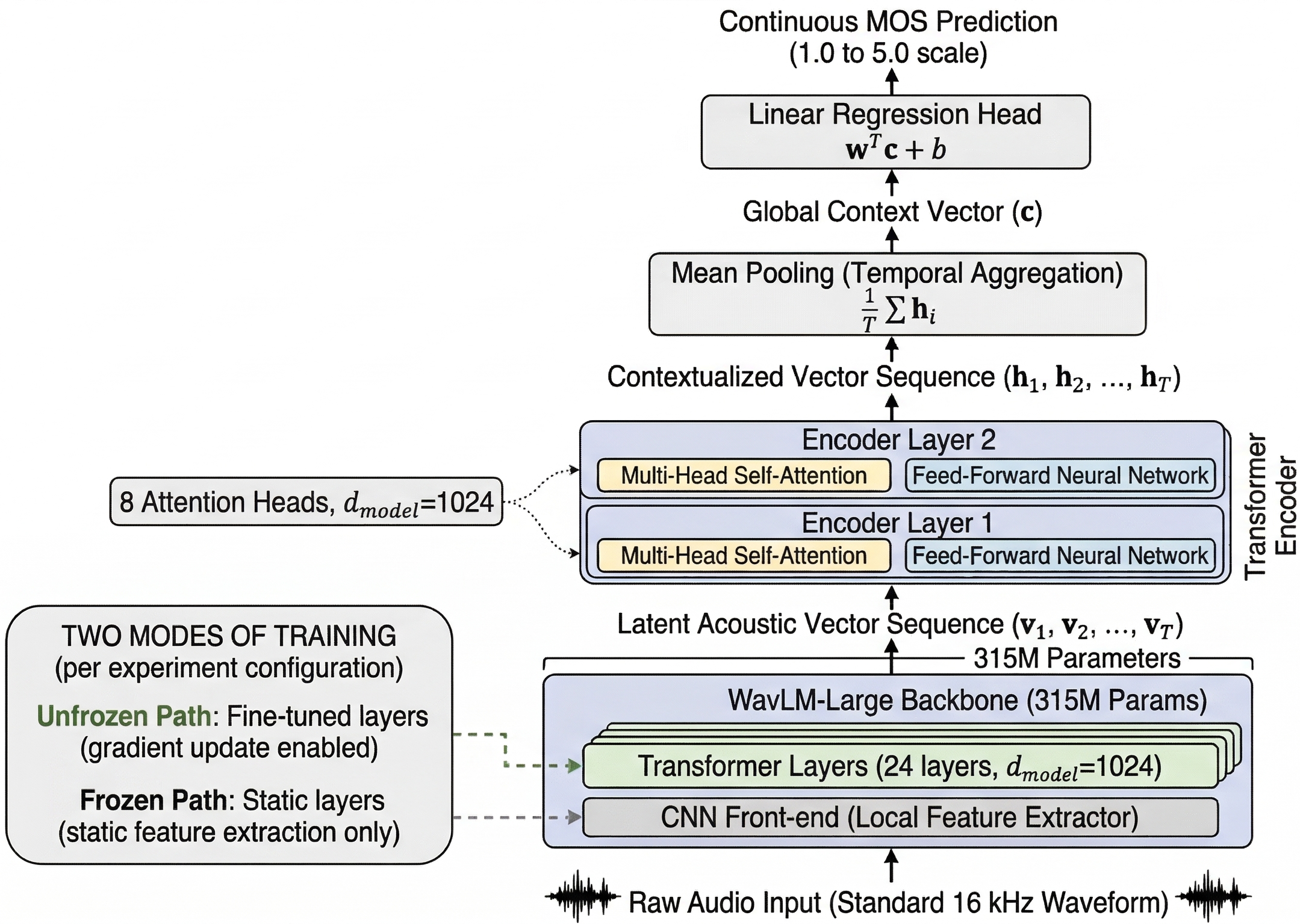}
	\caption{Architecture of the proposed SSL-Transformer Encoder framework.}
	\label{fig:ssl_transformer_arch}
\end{figure}

In this research, we implement a two-pronged strategy involving both frozen and fine-tuned versions of the WavLM backbone across the respective primary corpora for both parts of the study. The feature extraction stage utilizes the WavLM-Large backbone. For every audio input sampled at 16 kHz, WavLM generates a dense sequence of acoustic embeddings. Each vector in this sequence has a dimensionality of 1024, representing the acoustic information of roughly 20ms frames. For our primary performance benchmarks, the WavLM layers are fully unfrozen to allow the model’s 315 million parameters to refine their internal acoustic knowledge specifically for the nuances of human perceptual scales. However, for the specific out-of-dataset and unseen domain experiments conducted in both Part I and Part II, we utilize the frozen backbone as a static feature extractor. This strategy is adopted because full fine-tuning on such a massive scale is computationally prohibitive for every individual experimental case. Moreover, freezing the backbone allows us to evaluate the inherent generalization power of the pre-trained WavLM features, ensuring that the model’s performance on unseen datasets is a reflection of universal acoustic knowledge rather than task-specific memorization.

To manage the significant memory overhead of fine-tuning such a large architecture, we employ technical optimizations including Automatic Mixed Precision (AMP) and Gradient Checkpointing. These methods allow for stable training on 1024-dimensional feature spaces by optimizing numerical precision and memory allocation during the backward pass. This sequence of latent vectors is subsequently fed into a two-layer Transformer Encoder featuring 8-head self-attention. The primary function of this component is to perform bidirectional contextualization; through the Multi-Head Self-Attention mechanism, each frame in the sequence attends to every other frame regardless of their temporal distance. This process transforms the raw acoustic vectors into contextualized embeddings that account for global signal properties, such as the recency effect or the presence of transient artifacts like packet loss and clipping. Because the Transformer is a shape-preserving architecture, it outputs a sequence of the same length as the input, where each vector has been updated with information from the entire signal duration.

To convert this variable-length sequence into a fixed-size representation required for regression, the model employs Global Average Pooling (Mean Pooling) across the temporal dimension. This mathematical aggregation computes the average perceptual state of all frames, resulting in a single 1024-dimensional context vector that serves as a global summary of the utterance. This method was selected over the alternative CLS-token approach to ensure that every segment of the audio contributes equally to the final judgment, providing a more balanced assessment of long-duration signals. Finally, this context vector is passed to a Linear Regression layer with 1024 input nodes and a single output node, which maps the aggregated features to a continuous MOS prediction on a scale of 1.0 to 5.0. The architecture of the proposed SSL-Transformer Encoder framework was given in Figure \ref{fig:ssl_transformer_arch}.

\begin{figure}
	\centering
	\includegraphics[width=\textwidth]{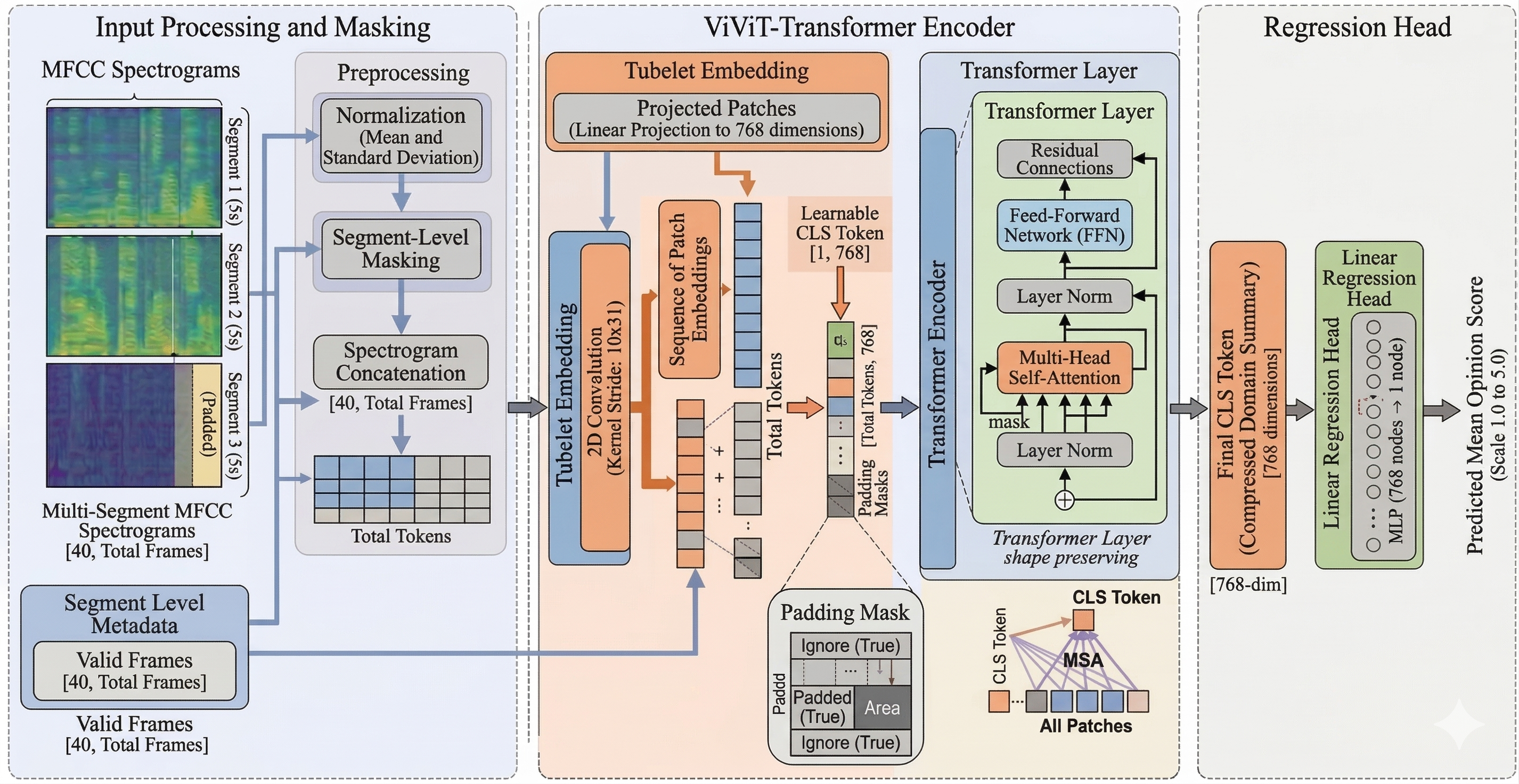}
	\caption{Architecture of the proposed ViViT-Transformer Encoder framework.}
	\label{fig:vivit_architecture}
\end{figure}

\subsubsection{ViViT-Transformer encoder architecture}

The second architecture proposed in this study utilizes a ViViT to process spectral data, treating the audio's time-frequency representation as a sequence of visual frames. In this framework, each audio signal is first decomposed into fixed-length segments of 5 seconds. To maintain architectural consistency across the dataset, zero-padding is applied selectively: it is only utilized when the final remaining segment of an utterance is shorter than the required 5-second window. This ensures that every MFCC matrix generated for a specific file has identical spatial dimensions, allowing them to be processed as a contiguous sequence of video frames. To prevent the model from learning from these artificial regions, we store specialized segment-level metadata. This information is used to generate a source key padding mask, which explicitly instructs the Transformer’s attention mechanism to ignore the zero-padded indices during both training and inference.

The processing pipeline begins with a Tubelet Embedding layer, implemented via a 2D convolution with a kernel and stride of $10 \times 31$. This layer partitions the 2D MFCC spectrogram into high-dimensional patches, which are then projected into a 768-dimensional embedding space. To represent the quality of the entire audio file, we prepend a single learnable CLS token in size 768 to this sequence of patch embeddings. While the MFCCs provide the raw spectral matter, the CLS token acts as a global observer. Because the Transformer Encoder is a shape-preserving architecture, every token—including the CLS token—remains 768-dimensional as it passes through the six layers of the encoder. Within each layer, the CLS token interacts with all spectral patches through Multi-Head Self-Attention, effectively aggregating the most salient quality features from the entire multi-segment sequence into its own vector representation.

The final Mean Opinion Score is derived by extracting the state of the CLS token from the output of the final Transformer layer. At this stage, the CLS token serves as a Compressed Domain Summary (CDS), containing a contextualized representation of the utterance's overall perceptual quality. This 768-dimensional summary is finally passed to a Linear Regression head (an multi layer perceptron with 768 input nodes and 1 output node), which maps the deep spectral features to a continuous MOS prediction on a scale of 1.0 to 5.0. This specialized file-level regression ensures that the model provides a single, unified quality score regardless of the original audio's duration. The architecture of the proposed ViViT-Transformer Encoder framework was given in Figure \ref{fig:vivit_architecture}.

\subsection{Experimental design and protocols}\label{subsec: experiments}

This research is structured into two primary parts to systematically evaluate the influence of data diversity and linguistic consistency on MOS prediction. By bifurcating the study, we aim to observe how the inclusion of non-speech vocalizations (singing) and multi-lingual samples affects the model's ability to generalize compared to a specialized, monolingual speech environment.

First part utilizes the entire consolidated dataset (130,652 Samples), including singing voice synthesis and multi-lingual samples (e.g., French and Japanese tracks from the VoiceMOS challenges). The objective of Part I is to assess the robustness of the SSL-Transformer and ViViT architectures when faced with high acoustic variance and out-of-distribution vocal signals that go beyond standard conversational English speech.

In second part, the corpus is refined to include only English-language speech (123,405 Samples). By removing singing and non-English data, we isolate the models' performance on a linguistically consistent domain. This allows for a direct comparison with Part I to determine if a narrower, language-specific focus improves the precision of quality assessment for traditional TTS and telephony applications.

In both parts of this study, we establish a performance baseline by training our models on the respective consolidated corpora to predict the MOS labels of the URGENT 2024 evaluation set. We selected URGENT 2024 as our primary benchmark because it represents a universal speech enhancement challenge, containing approximately 6,900 waveforms that cover a vast array of distortions, including additive noise, reverberation, bandwidth limitation, and clipping. Unlike more recent iterations of the challenge which may rely on model-based MOS predictions, URGENT 2024 provides high-quality ground-truth human labels, making it the most appropriate in-the-wild test for assessing how well our models bridge the gap between massive synthetic training data and real-world human perception \cite{URGENT2024}.

The architectural approach for this baseline involves a comprehensive comparison between the SSL-Transformer and ViViT-MFCC frameworks across both parts. For the SSL-Transformer model, we utilize both fine-tuned and frozen configurations to measure the specific performance gains achieved through end-to-end adaptation of the 315-million-parameter WavLM-Large backbone. Simultaneously, the ViViT-MFCC model is evaluated to determine the efficacy of processing spectral video patches for quality assessment in both broad and English-only contexts.

Part I of our experimental framework utilizes the full 130,652-sample corpus and focuses on the model's ability to generalize to entirely new datasets. For this phase, we selected five core datasets—VCC 2020 and the four VoiceMOS Challenges. These five datasets were chosen specifically because their sample sizes are statistically balanced, with each contributing approximately 20\% to this experimental subset. This balanced distribution ensures that no single data source dominates the training process, allowing for a fair and mathematically sound LODO evaluation.

We conducted six distinct experiments within this part. In the first five trials, one dataset was strictly isolated for testing, while the training and validation sets were generated randomly from the remaining four. In the sixth experiment, the model was trained and tested on a randomized mixture of all five datasets. By comparing the results of the five LODO cases against this sixth baseline, we can precisely quantify the performance degradation that occurs when the model encounters an unseen data distribution versus one it has previously observed. These experiments were conducted using the frozen SSL+Transformer and ViViT architectures. We opted for the frozen SSL configuration here because full fine-tuning on every individual case is computationally prohibitive, and freezing the backbone provides a more accurate measure of the pre-trained WavLM’s inherent generalization strength across diverse challenges.

The second part of our study replicates the protocols but restricts the scope to the 123,405-sample English-only corpus. This phase shifts the focus toward domain-specific robustness across highly distinct acoustic scenarios. In this part, the selected subset for experiments consisted of Blizzard 2011 (507 samples), VoiceMOS 2024 Tracks 3 (380 samples), TTSDS2 (460 samples), CHiME-7 (688 samples), and TCD-VoIP (384 samples). These datasets were selected because they have similar sizes, and also represent polar extremes of audio quality, ranging from domestic multi-speaker noise in CHiME-7 to high-fidelity neural text-to-speech in TTSDS2 and specialized VoIP-specific degradations in TCD-VoIP. The near-equal sample sizes across these highly distinct domains allow us to observe how domain-specific features—rather than sheer data volume—affect the accuracy of MOS predictions. By maintaining an English-only focus in this part, we eliminate cross-linguistic variables, allowing for a pure assessment of how the frozen SSL+Transformer and ViViT models handle varying levels of noise, synthetic artifacts, and communication channel distortions.

\subsection{Training Strategy and Computational Infrastructure}
\label{sec:training_strategy}
The training and benchmarking of the proposed architectures were executed across a dual-environment infrastructure designed to handle both heavy end-to-end backpropagation and high-throughput local evaluations. End-to-end fine-tuning of the $315$-million-parameter SSL-FT was conducted on cloud-hosted NVIDIA T4 GPUs ($16$\,GB VRAM). Due to gradient updates propagating through all 24 layers of WavLM-Large, a single epoch on the $130\text{k}$ broad corpus required approximately 9 to 10 hours. 

To ensure strict experimental control and direct hardware comparability for efficiency claims, all systematic LODO trials, feature extraction, and inference throughput benchmarks for both the ViViT-MFCC and SSL-FRZ architectures were conducted on a single local workstation equipped with an NVIDIA GeForce RTX 5070 Ti Laptop GPU ($12$\,GB VRAM), an Intel Core Ultra 9 processor, and $64$\,GB DDR5 RAM ($6000$\,MHz). On this unified local setup, the ViViT architecture completed a full training epoch in approximately 15 minutes under identical batch processing conditions, representing a $40\times$ speed advantage during downstream training compared to full SSL fine-tuning.

All models were optimized using the AdamW optimizer. A dual learning rate schedule was implemented: a conservative $10^{-5}$ for fine-tuning the pre-trained WavLM backbone to preserve its acoustic representations, and $10^{-4}$ for the frozen SSL downstream Transformer layers and ViViT models. Training stability was maintained using a multi-tiered checkpointing pipeline, and early stopping was enforced with a patience of 7 epochs monitored on validation MSE.

\subsection{Evaluation of results}

To provide a comprehensive assessment of our models, we evaluate their performance at two distinct granularities: utterance-level and system-level. This dual-evaluation approach ensures that we measure not only the accuracy of individual predictions but also the model's ability to rank and benchmark different synthesis and enhancement technologies.

For every experimental trial, we systematically record the predicted MOS and the ground-truth human MOS labels for each waveform in a structured CSV format. These raw results form the basis for our utterance-level analysis. To perform system-level evaluation, we aggregate the utterances within these CSV files based on their source dataset or synthesis system. By calculating the mean predicted MOS and comparing it against the mean human MOS for each group, we derive the system-level metrics, which are often more stable and reflective of the relative performance of different speech technologies.

The following statistical parameters are utilized across both evaluation granularities: MSE measures the average squared difference between predicted and actual scores. It serves as our primary indicator of absolute prediction accuracy. SRCC evaluates the monotonic relationship between predicted and true ratings. It is a critical metric for ranking, as it measures how well the model preserves the relative order of the samples or systems. LCC, also known as the Pearson correlation, measures the linear relationship between the two sets of scores, indicating the precision of the model’s linear mapping. KTAU is a robust rank-based metric that measures the ordinal association between rankings. It provides a reliable check for ranking stability, especially across diverse data distributions \cite{VOICEMOS2022}.

\subsection{Comparative Benchmarking via ARECHO}


\begin{table}[!htbp]
	\centering
	\footnotesize
	\caption{Comprehensive summary of SOTA metrics integrated into the ARECHO evaluation framework, categorized by type and range.}\label{tab:ARECHO_SOTA}
	
		
		\begin{tabular}{|l|l|l|l|l|l|}
			\hline
			\textbf{No.} & \textbf{Type} & \textbf{ARECHO Metric Name} & \textbf{Original Name} & \textbf{Range} & \textbf{Open Source Link} \\
			\hline
			1 & \multirow{10}{*}{Independent} & arecho\_dns\_overall & DNSMOS P.835 \cite{DNS_P835}& [1, 5] & \href{https://pypi.org/project/speechmos/}{Link} \\
			\cline{1-1} \cline{3-6}
			2 & & arecho\_dns\_p808 & DNSMOS P.808 \cite{DNS_P808}& [1, 5] & \href{https://pypi.org/project/speechmos/}{Link} \\
			\cline{1-1} \cline{3-6}
			3 & & arecho\_nisqa\_mos\_pred & NISQA (Overall MOS) \cite{NISQA}& [1, 5] & \href{https://github.com/gabrielmittag/NISQA}{Link} \\
			\cline{1-1} \cline{3-6}
			4 & & arecho\_plcmos & PLCMOS \cite{PLCMOS}& [1, 5] & \href{https://pypi.org/project/speechmos/}{Link} \\
			\cline{1-1} \cline{3-6}
			5 & & arecho\_scoreq\_nr & SCOREQ (No-Reference) \cite{SCOREQ}& [1, 5] & \href{https://github.com/ftshijt/scoreq/tree/main}{Link} \\
			\cline{1-1} \cline{3-6}
			6 & & arecho\_sheet\_ssqa & SSQA (SHEET Toolkit) \cite{SHEET}& [1, 5] & \href{https://github.com/unilight/sheet/tree/main}{Link} \\
			\cline{1-1} \cline{3-6}
			7 & & arecho\_utmos & UTMOS \cite{UTMOS}& [1, 5] & \href{https://github.com/tarepan/SpeechMOS}{Link} \\
			\cline{1-1} \cline{3-6}
			8 & & arecho\_utmosv2 & UTMOSv2 \cite{UTMOSv2}& [1, 5] & \href{https://github.com/sarulab-speech/UTMOSv2}{Link} \\
			\cline{1-1} \cline{3-6}
			9 & & arecho\_singmos & SingMOS \cite{SINGMOS}& [1, 5] & \href{https://github.com/South-Twilight/SingMOS}{Link} \\
			\cline{1-1} \cline{3-6}
			10 & & arecho\_audiobox\_aest\_PQ & Audiobox (Production Quality) \cite{MetaAudioboxAesthetics}& [1, 10] & \href{https://github.com/facebookresearch/audiobox-aesthetics}{Link} \\
			\hline
			11 & \multirow{2}{*}{Non-match} & arecho\_noresqa\_score & Noresqa (Non-matching Ref) \cite{NORESQA}& [1, 5] & \href{https://github.com/shimhz/Noresqa}{Link} \\
			\cline{1-1} \cline{3-6}
			12 & & arecho\_nomad & NOMAD (Non-matching Ref) \cite{NOMAD}& [1, 5] & \href{https://github.com/shimhz/nomad/tree/main}{Link} \\
			\hline
			13 & \multirow{3}{*}{Dependent} & arecho\_pesq & PESQ (Reference-based) \cite{PESQ}& [1, 5] & \href{https://pypi.org/project/pesq/}{Link} \\
			\cline{1-1} \cline{3-6}
			14 & & arecho\_visqol & ViSQOL (Reference-based) \cite{VISQOLv3}& [1, 5] & \href{https://github.com/google/visqol}{Link} \\
			\cline{1-1} \cline{3-6}
			15 & & arecho\_pysepm\_c\_ovl & PYSEPM (overall quality) \cite{PYSEPM}& [1, 5] & \href{https://github.com/shimhz/pysepm}{Link} \\
			\cline{1-1} \cline{3-6}
			\hline
			16 & \multirow{3}{*}{Ground-truth} & arecho\_urgent\_mos & URGENT MOS \cite{ARECHO}& [1, 5] & \href{https://github.com/wavlab-speech/versa/tree/main}{Link} \\
			\cline{1-1} \cline{3-6}
			17 & & arecho\_voicemos\_real\_mos & VoiceMOS Real MOS \cite{ARECHO}& [1, 5] & \href{https://github.com/wavlab-speech/versa/tree/main}{Link} \\
			\cline{1-1} \cline{3-6}
			18 & & arecho\_nisqa\_real\_mos & NISQA Real MOS \cite{ARECHO}& [1, 5] & \href{https://github.com/wavlab-speech/versa/tree/main}{Link} \\
			\hline
		\end{tabular}
\end{table}


To further ground our findings within the current SOTA, we utilize the ARECHO \cite{ARECHO} framework for multi-metric estimation. Unlike traditional MOS predictors that estimate quality in isolation, ARECHO employs a dynamic classifier chain to model the inter-metric dependencies across a diverse array of speech assessment dimensions.

\begin{figure*}[pos=t]
	\centering
	\begin{subfigure}[b]{0.46\textwidth}
		\centering
		\includegraphics[width=\linewidth]{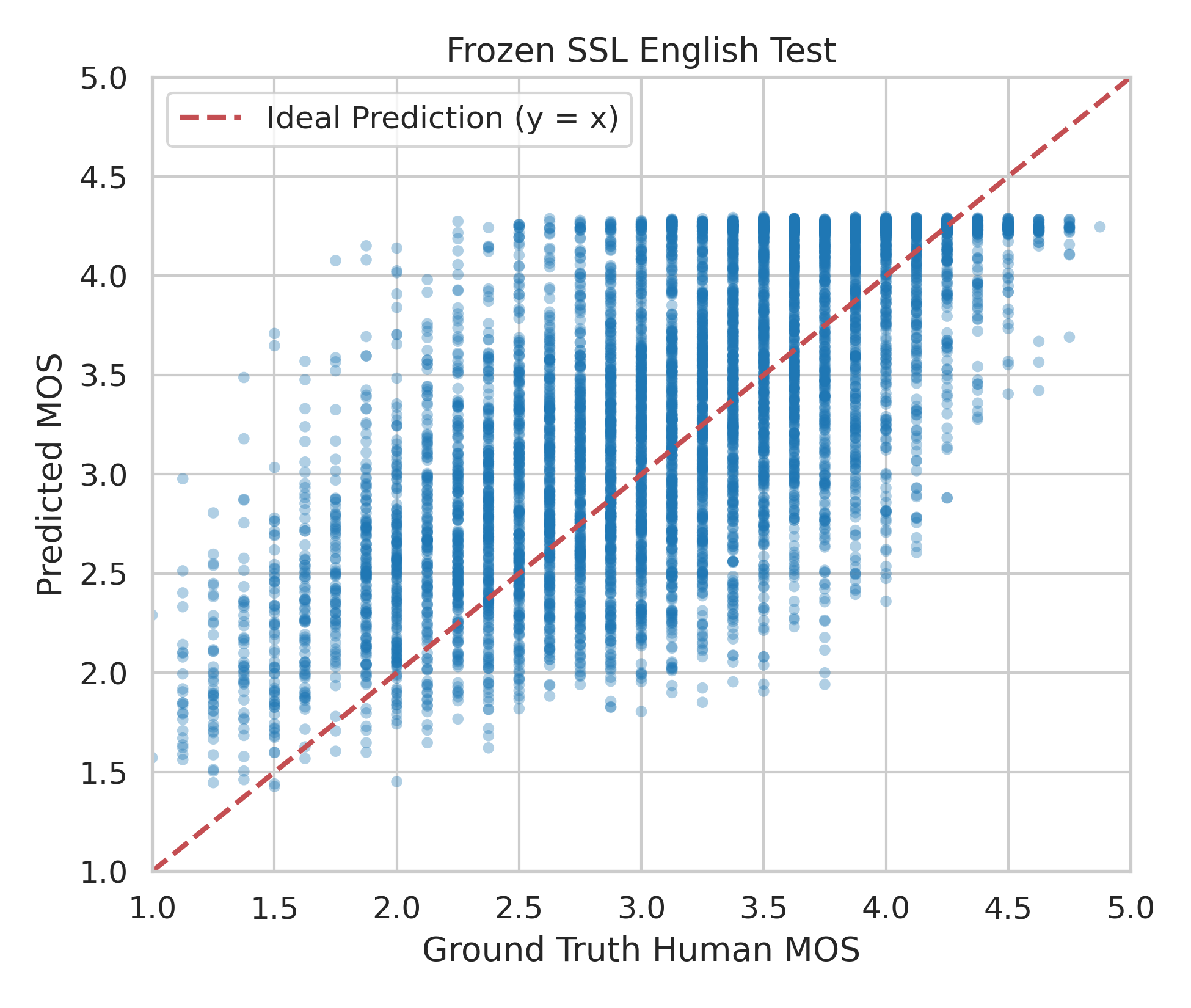}
		\caption{Frozen SSL (Test Set)}
		\label{fig:scatter_frozen_test}
	\end{subfigure}
	\hfill
	\begin{subfigure}[b]{0.46\textwidth}
		\centering
		\includegraphics[width=\linewidth]{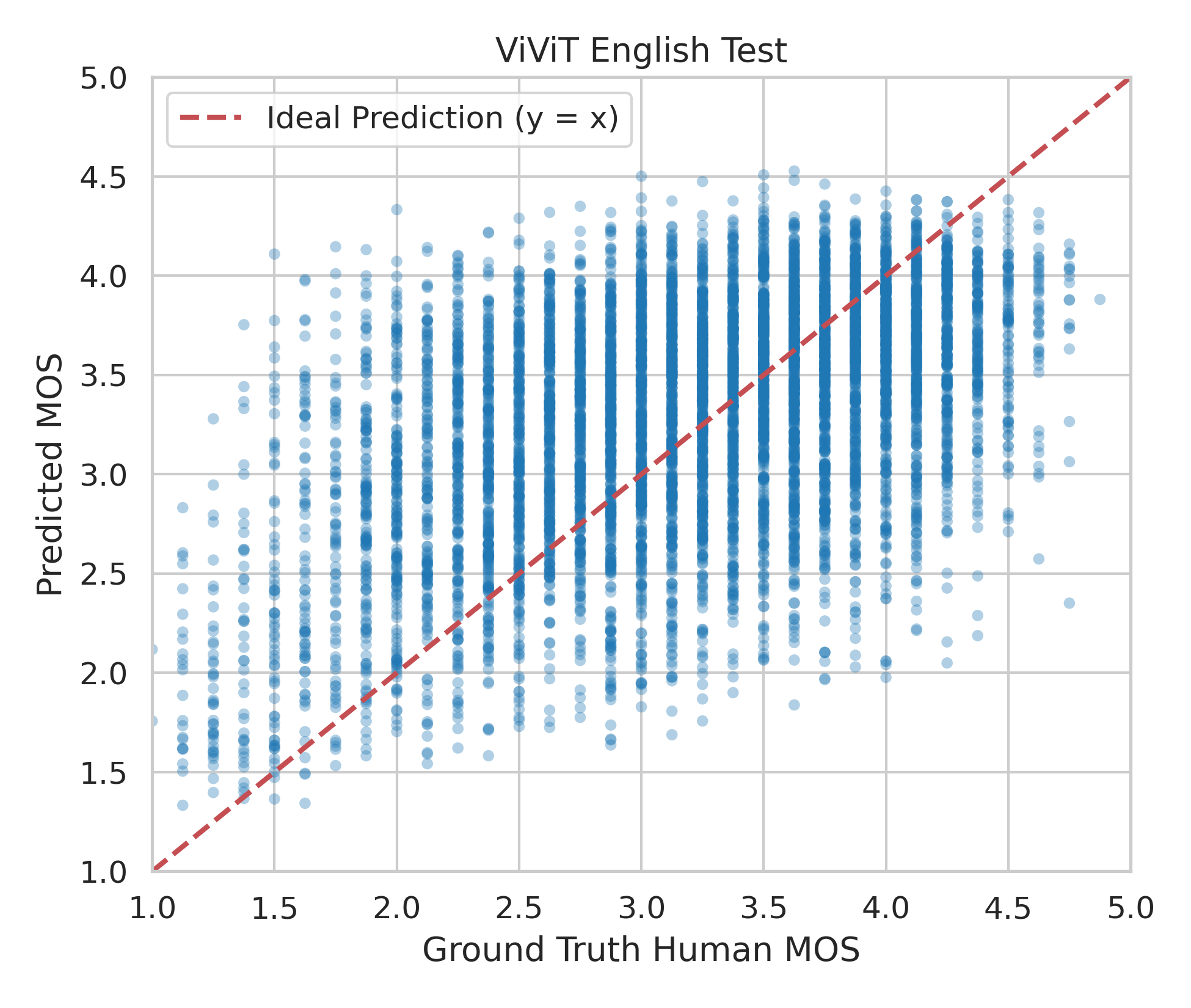}
		\caption{ViViT (Test Set)}
		\label{fig:scatter_vivit_test}
	\end{subfigure}
	
	\vspace{0.2cm}
	
	\begin{subfigure}[b]{0.46\textwidth}
		\centering
		\includegraphics[width=\linewidth]{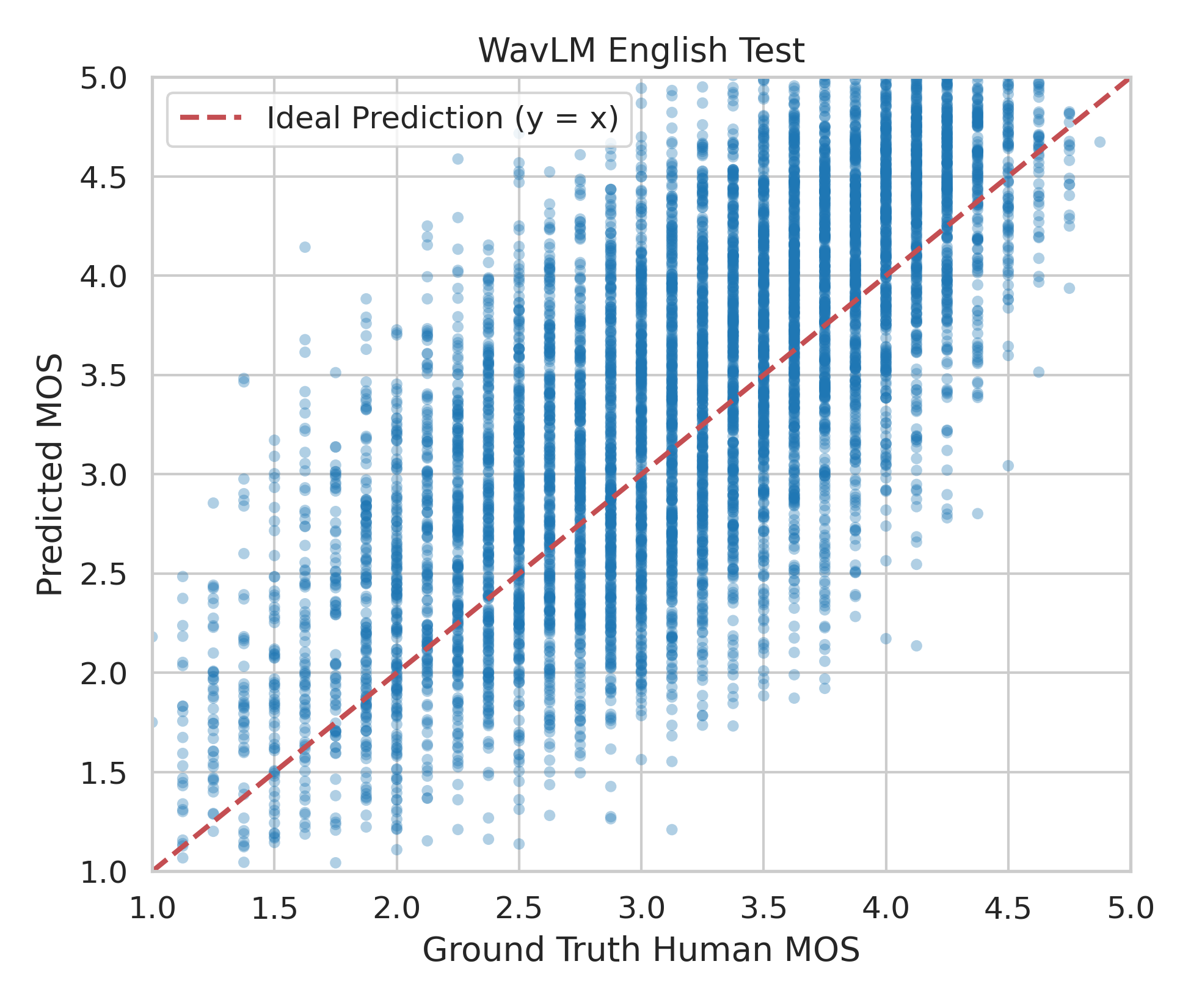}
		\caption{WavLM (Test Set)}
		\label{fig:scatter_wavlm_test}
	\end{subfigure}
	\hfill
	\begin{subfigure}[b]{0.46\textwidth}
		\centering
		\includegraphics[width=\linewidth]{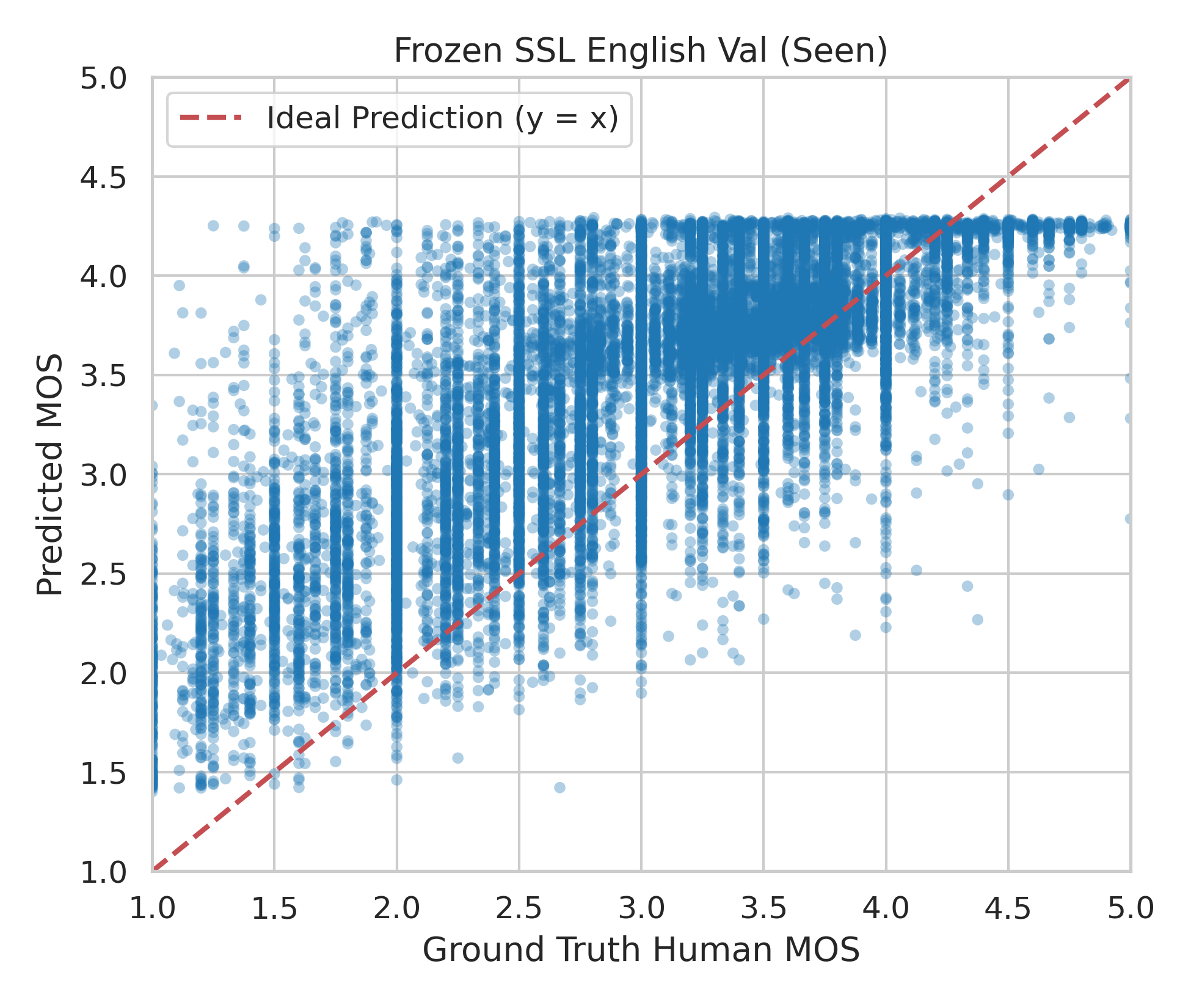}
		\caption{Frozen SSL (Validation Set)}
		\label{fig:scatter_frozen_val}
	\end{subfigure}
	
	\caption{Comparison of Ground Truth Human MOS vs. Predicted MOS across model architectures and evaluation splits.}
	\label{fig:all_scatter_plots}
\end{figure*}

\begin{figure}[pos=htbp]
	\centering
	\includegraphics[width=0.88\linewidth]{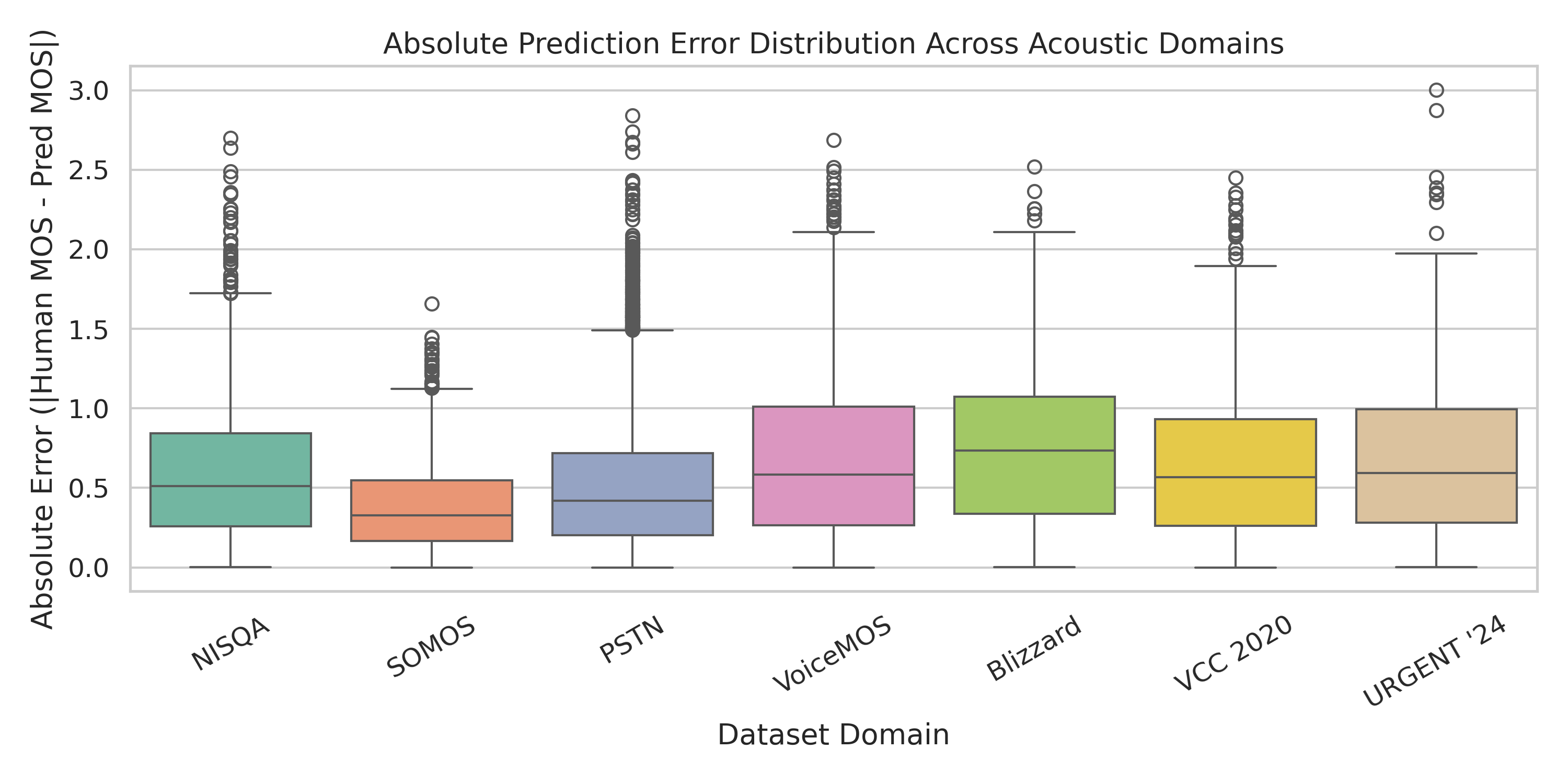}
	\caption{Absolute prediction error distribution across acoustic domains for the English corpus.}
	\label{fig:error_boxplots}
\end{figure}

A critical feature of our implementation is the use of ARECHO's non-intrusive inference mode. Although metrics like PESQ and ViSQOL traditionally require a clean reference signal, ARECHO can estimate these dependent metrics using only the processed test waveform. It achieves this by modeling the conditional probability of a metric given the latent speech tokens and the previously predicted metrics in the chain. This allows us to generate a full 18-metric quality profile (Table \ref{tab:ARECHO_SOTA}) without providing original reference audio, which is particularly useful for in-the-wild benchmarking where clean references are unavailable.

The benchmarks generated are categorized into four functional groups. Independent Metrics are naturally non-intrusive models, such as DNSMOS, UTMOS, and NISQA, which assess quality directly from the signal. Notably, since the arecho\_audiobox\_aest\_PQ operates on a 1–10 scale, we linearly normalized its output to a 1–5 range to ensure statistical consistency with the other integrated SOTA benchmarks during the comparative analysis. Dependent metrics represent metrics that typically require a time-aligned reference. In this study, ARECHO predicts these scores by observing signal degradations and estimating the distance from an implicit clean manifold. Unlike dependent metrics that require an exact temporal match, non-matching metrics are designed to compare a test signal against any high-quality reference from the same speaker, even if the spoken content is different. In our non-intrusive setup, ARECHO leverages its training on these models to estimate quality based on the consistency of the speaker's acoustic characteristics. Ground-truth estimations include URGENT MOS, VoiceMOS Real MOS, and NISQA Real MOS. Within the ARECHO framework, these function as oracle targets. By conditioning on other predicted acoustic features, the model estimates the likely human subjective score, providing a direct comparison to established seen human perception patterns.

\section{Results}\label{sec:result}

This section presents the empirical outcomes of the large-scale MOS prediction benchmarking. We provide the baseline performance metrics for the proposed architectures followed by the results of the systematic LODO protocols and the comprehensive SOTA benchmarking via ARECHO.

\subsection{Baseline Performance}


\begin{sidewaystable}[!htbp]
	\centering
	\footnotesize
	
	\caption{Comprehensive performance results for the proposed architectures. Models are evaluated across all individual datasets, categorized by their respective training phases (Part I and Part II). Metrics include MSE ($\downarrow$), LCC ($\uparrow$), SRCC ($\uparrow$), and KTAU ($\uparrow$).}
	\label{tab:main_corpus_results}
	
	
	\resizebox{\textwidth}{!}{
		
		\begin{tabular}{l | cccc | cccc | cccc || cccc | cccc | cccc}
			\toprule
			& \multicolumn{12}{c||}{\textbf{Part I Training (Broad Corpus Models)}} & \multicolumn{12}{c}{\textbf{Part II Training (English-Only Models)}} \\
			\cmidrule(lr){2-13} \cmidrule(lr){14-25}
			& \multicolumn{4}{c|}{SSL (FRZ)} & \multicolumn{4}{c|}{SSL (FT)} & \multicolumn{4}{c||}{ViViT} & \multicolumn{4}{c|}{SSL (FRZ)} & \multicolumn{4}{c|}{SSL (FT)} & \multicolumn{4}{c}{ViViT} \\
			\cmidrule(lr){2-5} \cmidrule(lr){6-9} \cmidrule(lr){10-13} \cmidrule(lr){14-17} \cmidrule(lr){18-21} \cmidrule(lr){22-25}
			\textbf{Dataset} & \tiny{MSE} & \tiny{LCC} & \tiny{SRCC} & \tiny{KTAU} & \tiny{MSE} & \tiny{LCC} & \tiny{SRCC} & \tiny{KTAU} & \tiny{MSE} & \tiny{LCC} & \tiny{SRCC} & \tiny{KTAU} & \tiny{MSE} & \tiny{LCC} & \tiny{SRCC} & \tiny{KTAU} & \tiny{MSE} & \tiny{LCC} & \tiny{SRCC} & \tiny{KTAU} & \tiny{MSE} & \tiny{LCC} & \tiny{SRCC} & \tiny{KTAU} \\
			\midrule
			Blizzard 2008 & 0.62 & 0.67 & 0.64 & 0.49 & \textbf{0.24} & \textbf{0.77} & \textbf{0.74} & \textbf{0.58} & 0.51 & 0.45 & 0.35 & 0.26 & 0.75 & 0.57 & 0.59 & 0.44 & 0.26 & 0.74 & 0.71 & 0.56 & 0.51 & 0.41 & 0.33 & 0.24 \\
			Blizzard 2009 & 0.62 & 0.76 & 0.68 & 0.53 & \textbf{0.21} & \textbf{0.85} & \textbf{0.79} & \textbf{0.64} & 0.43 & 0.62 & 0.47 & 0.36 & 0.60 & 0.72 & 0.66 & 0.51 & 0.23 & 0.84 & 0.78 & 0.63 & 0.45 & 0.57 & 0.40 & 0.29 \\
			Blizzard 2010 & 0.88 & 0.73 & 0.71 & 0.53 & 0.28 & \textbf{0.85} & \textbf{0.81} & \textbf{0.62} & 1.02 & 0.43 & 0.40 & 0.28 & 0.95 & 0.67 & 0.70 & 0.51 & \textbf{0.28} & 0.84 & 0.80 & 0.62 & 0.83 & 0.40 & 0.36 & 0.25 \\
			Blizzard 2011 & 0.64 & 0.79 & 0.76 & 0.56 & 0.17 & 0.84 & 0.83 & \textbf{0.67} & 0.61 & 0.44 & 0.38 & 0.26 & 1.09 & 0.77 & 0.81 & 0.61 & \textbf{0.16} & \textbf{0.86} & \textbf{0.83} & 0.67 & 0.47 & 0.44 & 0.37 & 0.28 \\
			Blizzard 2012 & 0.80 & 0.85 & 0.85 & 0.68 & 0.11 & 0.94 & 0.92 & 0.76 & 1.52 & 0.31 & 0.33 & 0.23 & 0.69 & 0.86 & 0.87 & 0.70 & \textbf{0.10} & \textbf{0.95} & \textbf{0.94} & \textbf{0.79} & 1.25 & 0.11 & 0.08 & 0.04 \\
			Blizzard 2013 & 0.63 & 0.86 & 0.87 & 0.69 & 0.16 & 0.91 & 0.89 & 0.72 & 0.78 & 0.59 & 0.58 & 0.41 & 0.87 & 0.82 & 0.81 & 0.63 & \textbf{0.16} & \textbf{0.91} & \textbf{0.90} & \textbf{0.73} & 0.63 & 0.56 & 0.58 & 0.39 \\
			Blizzard 2013 Ext. & 0.40 & 0.84 & 0.80 & \textbf{0.63} & 0.15 & 0.88 & 0.82 & 0.62 & 0.54 & 0.53 & 0.39 & 0.28 & 0.44 & 0.78 & 0.82 & 0.62 & \textbf{0.14} & \textbf{0.90} & \textbf{0.83} & \textbf{0.63} & 0.69 & 0.51 & 0.48 & 0.36 \\
			NISQA & 0.68 & 0.86 & 0.87 & 0.69 & \textbf{0.23} & \textbf{0.92} & \textbf{0.92} & \textbf{0.77} & 0.68 & 0.68 & 0.68 & 0.50 & 0.55 & 0.84 & 0.86 & 0.68 & 0.24 & 0.92 & 0.92 & 0.77 & 0.63 & 0.69 & 0.68 & 0.50 \\
			PSTN & 0.46 & 0.80 & 0.80 & 0.61 & 0.30 & 0.82 & 0.82 & 0.63 & 0.37 & 0.73 & 0.73 & 0.54 & 0.39 & 0.79 & 0.80 & 0.61 & \textbf{0.29} & \textbf{0.82} & \textbf{0.82} & \textbf{0.64} & 0.35 & 0.73 & 0.73 & 0.55 \\
			SOMOS & 0.18 & 0.54 & 0.52 & 0.36 & 0.09 & 0.62 & 0.60 & 0.43 & 0.15 & 0.17 & 0.17 & 0.12 & 0.22 & 0.46 & 0.45 & 0.32 & \textbf{0.09} & \textbf{0.63} & \textbf{0.61} & \textbf{0.44} & 0.15 & 0.20 & 0.21 & 0.15 \\
			VCC 2020 & 0.89 & 0.82 & 0.84 & 0.65 & 0.23 & 0.90 & 0.91 & 0.74 & 0.96 & 0.57 & 0.56 & 0.40 & 0.67 & 0.81 & 0.83 & 0.64 & \textbf{0.22} & \textbf{0.91} & \textbf{0.91} & \textbf{0.75} & 0.70 & 0.62 & 0.61 & 0.44 \\
			VoiceMOS 2022 & 0.85 & 0.68 & 0.66 & 0.49 & \textbf{0.27} & 0.83 & 0.82 & 0.64 & 0.79 & 0.49 & 0.48 & 0.34 & 0.74 & 0.70 & 0.71 & 0.52 & 0.28 & \textbf{0.83} & \textbf{0.83} & \textbf{0.64} & 0.59 & 0.49 & 0.48 & 0.34 \\
			VoiceMOS 2023 (T2) & 0.76 & 0.69 & 0.68 & 0.48 & \textbf{0.25} & \textbf{0.86} & \textbf{0.84} & \textbf{0.65} & 0.74 & 0.54 & 0.51 & 0.35 & --- & --- & --- & --- & --- & --- & --- & --- & --- & --- & --- & --- \\
			VoiceMOS 2024 (T2) & 0.62 & 0.48 & 0.50 & 0.36 & \textbf{0.43} & \textbf{0.64} & \textbf{0.62} & \textbf{0.46} & 0.59 & 0.41 & 0.43 & 0.31 & --- & --- & --- & --- & --- & --- & --- & --- & --- & --- & --- & --- \\
			VoiceMOS 2024 (T3) & 1.19 & 0.46 & 0.51 & 0.36 & 0.41 & 0.84 & 0.84 & 0.67 & 1.00 & 0.50 & 0.47 & 0.34 & 1.18 & 0.41 & 0.37 & 0.27 & \textbf{0.38} & \textbf{0.85} & \textbf{0.85} & \textbf{0.68} & 1.40 & 0.37 & 0.33 & 0.22 \\
			VoiceMOS 2025 (TTS) & 0.66 & 0.75 & 0.77 & 0.57 & \textbf{0.11} & 0.90 & \textbf{0.87} & 0.69 & 0.55 & 0.57 & 0.47 & 0.34 & 0.71 & 0.62 & 0.57 & 0.40 & 0.12 & \textbf{0.91} & 0.86 & \textbf{0.69} & 0.45 & 0.48 & 0.37 & 0.26 \\
			TTSDS2 & 1.15 & 0.32 & 0.28 & 0.18 & \textbf{0.37} & \textbf{0.54} & \textbf{0.44} & \textbf{0.32} & 0.86 & -0.04 & -0.03 & -0.02 & 1.30 & 0.23 & 0.21 & 0.14 & 0.42 & 0.47 & 0.36 & 0.26 & 0.73 & -0.20 & -0.13 & -0.09 \\
			CHiME-7 & 0.41 & 0.79 & 0.79 & 0.60 & 0.43 & 0.88 & 0.87 & 0.70 & 0.57 & 0.65 & 0.63 & 0.46 & \textbf{0.39} & 0.74 & 0.73 & 0.54 & 0.45 & \textbf{0.89} & \textbf{0.89} & \textbf{0.72} & 0.51 & 0.64 & 0.62 & 0.45 \\
			TCD-VoIP & 1.61 & 0.55 & 0.55 & 0.40 & 0.22 & 0.93 & \textbf{0.94} & \textbf{0.79} & 0.76 & 0.63 & 0.72 & 0.50 & 1.53 & 0.40 & 0.44 & 0.32 & \textbf{0.20} & \textbf{0.94} & 0.92 & 0.75 & 0.72 & 0.59 & 0.59 & 0.40 \\
			\midrule
			\textbf{Avg. Val Data} & 0.52 & 0.78 & 0.76 & 0.58 & 0.25 & \textbf{0.85} & \textbf{0.84} & \textbf{0.66} & 0.47 & 0.66 & 0.64 & 0.47 & 0.45 & 0.78 & 0.76 & 0.58 & \textbf{0.24} & \textbf{0.85} & \textbf{0.84} & \textbf{0.66} & 0.41 & 0.68 & 0.66 & 0.48 \\
			\midrule
			\textbf{URGENT 2024} & 0.43 & 0.71 & \textbf{0.72} & \textbf{0.53} & 0.44 & \textbf{0.72} & \textbf{0.72} & \textbf{0.53} & 0.82 & 0.44 & 0.39 & 0.28 & \textbf{0.36} & 0.71 & 0.71 & 0.52 & 0.45 & \textbf{0.72} & 0.71 & \textbf{0.53} & 0.50 & 0.48 & 0.44 & 0.31 \\
			\bottomrule
		\end{tabular}
	}
	
\end{sidewaystable}


The baseline results for the proposed architectures across both consolidated corpora are presented in Table \ref{tab:main_corpus_results}. In this comparison, the architectures are identified as SSL(FRZ) for the frozen self-supervised model, SSL(FT) for the fine-tuned version, and ViViT. For both Part I and Part II, the trained models were utilized to generate predictions for their respective validation sets as well as the URGENT 2024 test dataset to assess cross-domain capability.

The table is structured to provide granular performance for each individual dataset—19 validation sets for the Broad Corpus (Part I) and 17 for the English-Only Corpus (Part II). These individual dataset results constitute the system-level performance metrics. To assess the utterance-level performance across the broader corpus, the table provides the average validation results and the results for the URGENT 2024 test dataset at the bottom of each section. Performance is quantified using four standard metrics: MSE for absolute error, and LCC, SRCC, and KTAU for ranking and correlation stability. To facilitate immediate comparison, the highest-performing metric for each individual dataset and the aggregate averages have been highlighted in bold. 

Across the validation datasets in both Part I and Part II, the SSL(FT) architecture consistently recorded the highest performance across all four metrics. Conversely, the ViViT architecture recorded the lowest performance ceiling among the three models; however, its results remained within a comparable range to the SSL-based trials. Notably, the performance trends for ViViT followed a similar trajectory to the SSL models when transitioning between corpora, showing a measurable improvement in both average validation and test metrics in the English-only (Part II) case compared to the Broad Corpus (Part I).

On the URGENT 2024 test dataset specifically, the performance rankings shifted. In the English-only corpus, the SSL(FRZ) model achieved the lowest MSE of 0.36, representing the leading absolute prediction accuracy for this unseen test set. This was accompanied by superior results in the correlation metrics (LCC, SRCC, and KTAU) for the SSL(FRZ) trial, contrasting with the rankings observed on the seen validation data where SSL(FT) was dominant.

As illustrated in Figure \ref{fig:all_scatter_plots}, the scatter plots compare human ground-truth MOS against predicted MOS across the architectures. Specifically, the fine-tuned WavLM model (Figure \ref{fig:scatter_wavlm_test}) and frozen SSL model (Figures \ref{fig:scatter_frozen_test} and \ref{fig:scatter_frozen_val}) demonstrate tight alignment along the diagonal, whereas the ViViT model (Figure \ref{fig:scatter_vivit_test}) exhibits wider variance.

\subsection{LODO experiments results}

To measure the impact of domain shift, we report the results of the 6-case systematic LODO experiment for both parts.


\begin{table}[!htbp]
	\centering
	\footnotesize
	
	\caption{Systematic 6-case LODO results for Part I (Broad Corpus). Performance metrics for Frozen SSL and ViViT architectures are grouped by experimental Case. For each evaluation target, \textit{Val} represents in-domain validation and \textit{Test} represents out-of-domain cross-corpus performance.}
	\label{tab:lodo_exp_results_part_1}
	
	
	\resizebox{\textwidth}{!}{
		
		\begin{tabular}{llc | cc | cc | cc | cc | cc || cc}
			\toprule
			& & & \multicolumn{10}{c|}{\textbf{Individual Evaluation Targets}} & \multicolumn{2}{c}{\textbf{System Summary}} \\
			\cmidrule(lr){4-13} \cmidrule(lr){14-15}
			& & & \multicolumn{2}{c|}{\textbf{VCC '20}} & \multicolumn{2}{c|}{\textbf{VMOS '22}} & \multicolumn{2}{c|}{\textbf{VMOS '23}} & \multicolumn{2}{c|}{\textbf{VMOS '24}} & \multicolumn{2}{c||}{\textbf{VMOS '25}} & \textbf{System} & \textbf{System} \\
			\textbf{Case} & \textbf{Model} & \textbf{Stat} & Val & Test & Val & Test & Val & Test & Val & Test & Val & Test & \textbf{Avg. Val} & \textbf{Avg. Test} \\
			\midrule
			
			\multirow{8}{*}{\textbf{\shortstack[l]{Case 1:\\VCC}}} 
			& \multirow{4}{*}{SSL (FRZ)} & MSE  & ---  & \textbf{0.54} & 0.48 & ---  & 0.50 & ---  & 0.48 & ---  & \textbf{0.26} & ---  & 0.46 & 0.54 \\
			&                           & LCC  & ---  & \textbf{0.83} & 0.77 & ---  & 0.74 & ---  & 0.59 & ---  & \textbf{0.85} & ---  & 0.79 & \textbf{0.83} \\
			&                           & SRCC & ---  & \textbf{0.83} & 0.77 & ---  & 0.70 & ---  & 0.57 & ---  & \textbf{0.84} & ---  & 0.78 & \textbf{0.83} \\
			&                           & KTAU & ---  & \textbf{0.64} & 0.58 & ---  & 0.51 & ---  & 0.42 & ---  & \textbf{0.66} & ---  & 0.59 & \textbf{0.64} \\
			\cmidrule{2-15}
			& \multirow{4}{*}{ViViT}     & MSE  & ---  & 1.18 & 0.65 & ---  & 0.74 & ---  & 0.62 & ---  & 0.32 & ---  & 0.63 & 1.18 \\
			&                           & LCC  & ---  & 0.17 & 0.39 & ---  & 0.42 & ---  & 0.40 & ---  & 0.62 & ---  & 0.52 & 0.17 \\
			&                           & SRCC & ---  & 0.15 & 0.37 & ---  & 0.40 & ---  & 0.37 & ---  & 0.58 & ---  & 0.51 & 0.15 \\
			&                           & KTAU & ---  & 0.10 & 0.26 & ---  & 0.28 & ---  & 0.27 & ---  & 0.42 & ---  & 0.36 & 0.10 \\
			\midrule
			
			\multirow{8}{*}{\textbf{\shortstack[l]{Case 2:\\VM22}}} 
			& \multirow{4}{*}{SSL (FRZ)} & MSE  & 0.41 & ---  & ---  & \textbf{0.64} & 0.59 & ---  & 0.68 & ---  & \textbf{0.18} & ---  & 0.48 & 0.64 \\
			&                           & LCC  & \textbf{0.85} & ---  & ---  & \textbf{0.59} & 0.66 & ---  & 0.59 & ---  & 0.83 & ---  & 0.78 & 0.59 \\
			&                           & SRCC & \textbf{0.86} & ---  & ---  & \textbf{0.58} & 0.64 & ---  & 0.56 & ---  & 0.74 & ---  & 0.75 & 0.58 \\
			&                           & KTAU & \textbf{0.67} & ---  & ---  & \textbf{0.41} & 0.47 & ---  & 0.41 & ---  & 0.56 & ---  & 0.57 & 0.41 \\
			\cmidrule{2-15}
			& \multirow{4}{*}{ViViT}     & MSE  & 0.53 & ---  & ---  & 0.89 & 0.64 & ---  & 0.56 & ---  & \textbf{0.18} & ---  & 0.51 & 0.89 \\
			&                           & LCC  & 0.72 & ---  & ---  & 0.23 & 0.54 & ---  & 0.55 & ---  & 0.82 & ---  & 0.70 & 0.23 \\
			&                           & SRCC & 0.71 & ---  & ---  & 0.20 & 0.49 & ---  & 0.51 & ---  & 0.74 & ---  & 0.68 & 0.20 \\
			&                           & KTAU & 0.53 & ---  & ---  & 0.14 & 0.35 & ---  & 0.37 & ---  & 0.55 & ---  & 0.50 & 0.14 \\
			\midrule
			
			\multirow{8}{*}{\textbf{\shortstack[l]{Case 3:\\VM23}}} 
			& \multirow{4}{*}{SSL (FRZ)} & MSE  & 0.40 & ---  & 0.41 & ---  & ---  & 2.53 & 0.68 & ---  & 0.28 & ---  & \textbf{0.45} & 2.53 \\
			&                           & LCC  & \textbf{0.85} & ---  & 0.75 & ---  & ---  & \textbf{0.34} & 0.55 & ---  & 0.75 & ---  & 0.80 & 0.34 \\
			&                           & SRCC & \textbf{0.85} & ---  & 0.74 & ---  & ---  & \textbf{0.34} & 0.48 & ---  & 0.69 & ---  & 0.78 & 0.34 \\
			&                           & KTAU & \textbf{0.66} & ---  & 0.55 & ---  & ---  & \textbf{0.24} & 0.35 & ---  & 0.50 & ---  & 0.59 & 0.24 \\
			\cmidrule{2-15}
			& \multirow{4}{*}{ViViT}     & MSE  & 0.60 & ---  & 0.54 & ---  & ---  & \textbf{1.32} & 0.58 & ---  & \textbf{0.27} & ---  & 0.54 & 1.32 \\
			&                           & LCC  & 0.69 & ---  & 0.57 & ---  & ---  & 0.22 & 0.46 & ---  & 0.75 & ---  & 0.68 & 0.22 \\
			&                           & SRCC & 0.68 & ---  & 0.56 & ---  & ---  & 0.21 & 0.47 & ---  & 0.69 & ---  & 0.67 & 0.21 \\
			&                           & KTAU & 0.50 & ---  & 0.40 & ---  & ---  & 0.14 & 0.33 & ---  & 0.51 & ---  & 0.49 & 0.14 \\
			\midrule
			
			\multirow{8}{*}{\textbf{\shortstack[l]{Case 4:\\VM24}}} 
			& \multirow{4}{*}{SSL (FRZ)} & MSE  & 0.45 & ---  & 0.43 & ---  & 0.54 & ---  & ---  & \textbf{1.08} & 0.29 & ---  & \textbf{0.45} & 1.08 \\
			&                           & LCC  & \textbf{0.82} & ---  & 0.71 & ---  & 0.68 & ---  & ---  & \textbf{0.40} & 0.74 & ---  & 0.76 & 0.40 \\
			&                           & SRCC & \textbf{0.85} & ---  & 0.71 & ---  & 0.66 & ---  & ---  & \textbf{0.36} & 0.76 & ---  & 0.76 & 0.36 \\
			&                           & KTAU & \textbf{0.66} & ---  & 0.53 & ---  & 0.48 & ---  & ---  & \textbf{0.26} & 0.57 & ---  & 0.57 & 0.26 \\
			\cmidrule{2-15}
			& \multirow{4}{*}{ViViT}     & MSE  & 0.54 & ---  & 0.53 & ---  & 0.59 & ---  & ---  & 1.78 & \textbf{0.23} & ---  & 0.51 & 1.78 \\
			&                           & LCC  & 0.73 & ---  & 0.58 & ---  & 0.58 & ---  & ---  & 0.17 & 0.78 & ---  & 0.67 & 0.17 \\
			&                           & SRCC & 0.73 & ---  & 0.56 & ---  & 0.54 & ---  & ---  & 0.16 & 0.67 & ---  & 0.66 & 0.16 \\
			&                           & KTAU & 0.54 & ---  & 0.40 & ---  & 0.38 & ---  & ---  & 0.11 & 0.50 & ---  & 0.48 & 0.11 \\
			\midrule
			
			\multirow{8}{*}{\textbf{\shortstack[l]{Case 5:\\VM25}}} 
			& \multirow{4}{*}{SSL (FRZ)} & MSE  & 0.63 & ---  & 0.64 & ---  & 0.62 & ---  & 0.54 & ---  & ---  & 0.89 & 0.61 & 0.89 \\
			&                           & LCC  & \textbf{0.86} & ---  & 0.78 & ---  & 0.73 & ---  & 0.64 & ---  & ---  & \textbf{0.48} & \textbf{0.80} & 0.48 \\
			&                           & SRCC & \textbf{0.87} & ---  & 0.78 & ---  & 0.69 & ---  & 0.62 & ---  & ---  & \textbf{0.42} & \textbf{0.79} & 0.42 \\
			&                           & KTAU & \textbf{0.69} & ---  & 0.59 & ---  & 0.51 & ---  & 0.46 & ---  & ---  & \textbf{0.28} & \textbf{0.61} & 0.28 \\
			\cmidrule{2-15}
			& \multirow{4}{*}{ViViT}     & MSE  & 0.70 & ---  & \textbf{0.50} & ---  & 0.62 & ---  & 0.58 & ---  & ---  & \textbf{0.86} & 0.59 & 0.86 \\
			&                           & LCC  & 0.62 & ---  & 0.58 & ---  & 0.52 & ---  & 0.53 & ---  & ---  & 0.10 & 0.63 & 0.10 \\
			&                           & SRCC & 0.61 & ---  & 0.57 & ---  & 0.49 & ---  & 0.50 & ---  & ---  & 0.07 & 0.62 & 0.07 \\
			&                           & KTAU & 0.45 & ---  & 0.41 & ---  & 0.34 & ---  & 0.36 & ---  & ---  & 0.04 & 0.45 & 0.04 \\
			\midrule
			
			\multirow{8}{*}{\textbf{\shortstack[l]{Case 6:\\Mixed}}} 
			& \multirow{4}{*}{SSL (FRZ)} & MSE  & 0.51 & 0.50 & 0.48 & 0.49 & 0.67 & 0.53 & 0.53 & 0.53 & 0.26 & 0.31 & 0.51 & \textbf{0.49} \\
			&                           & LCC  & \textbf{0.82} & \textbf{0.79} & 0.70 & 0.69 & 0.62 & 0.69 & 0.57 & 0.58 & 0.82 & 0.78 & 0.75 & 0.75 \\
			&                           & SRCC & \textbf{0.83} & \textbf{0.81} & 0.70 & 0.68 & 0.58 & 0.65 & 0.53 & 0.55 & 0.75 & 0.70 & 0.76 & 0.75 \\
			&                           & KTAU & \textbf{0.65} & \textbf{0.62} & 0.52 & 0.50 & 0.41 & 0.47 & 0.38 & 0.40 & 0.56 & 0.52 & 0.57 & 0.56 \\
			\cmidrule{2-15}
			& \multirow{4}{*}{ViViT}     & MSE  & 0.56 & 0.56 & 0.56 & 0.56 & 0.74 & 0.73 & 0.61 & 0.55 & \textbf{0.20} & \textbf{0.23} & 0.57 & 0.56 \\
			&                           & LCC  & 0.71 & 0.69 & 0.56 & 0.56 & 0.50 & 0.51 & 0.45 & 0.53 & 0.79 & 0.77 & 0.66 & 0.65 \\
			&                           & SRCC & 0.71 & 0.68 & 0.55 & 0.55 & 0.45 & 0.46 & 0.42 & 0.48 & 0.72 & 0.68 & 0.65 & 0.64 \\
			&                           & KTAU & 0.52 & 0.50 & 0.39 & 0.39 & 0.32 & 0.32 & 0.30 & 0.34 & 0.55 & 0.50 & 0.47 & 0.46 \\
			\bottomrule
		\end{tabular}
	}
\end{table}


\subsubsection{LODO experiment results of part I}

Table \ref{tab:lodo_exp_results_part_1} details the generalization capabilities of the SSL(FRZ) and ViViT architectures when a single dataset is completely isolated from the training process. This systematic 6-case experiment evaluates zero-shot generalization on isolated unseen datasets (cases 1–5) and establishes a seen-distribution benchmark using a randomized mixture of data (case 6).

The table compares the performance of the two architectures across the five individual evaluation datasets, with separate reporting for validation and test results. For each dataset and specific experimental case, the leading metric values between the two models are highlighted in bold for both test and validation. In the final two columns, representing the System Summary, a vertical comparison is conducted across all cases and models to identify and bold the overall best performance for the aggregate validation and test benchmarks.

\begin{table}[!ht]
	\centering
	\footnotesize
	\caption{Systematic 6-case LODO results for Part II (Narrow Corpus). Performance metrics for Frozen SSL and ViViT architectures are grouped by experimental Case. For each evaluation target, \textit{Val} represents in-domain validation and \textit{Test} represents out-of-domain cross-corpus performance.}
	\label{tab:lodo_exp_results_part_2}
	
	
	\resizebox{\textwidth}{!}{
		\begin{tabular}{llc | cc | cc | cc | cc | cc || cc}
			\toprule
			& & & \multicolumn{10}{c|}{\textbf{Individual Evaluation Targets}} & \multicolumn{2}{c}{\textbf{System Summary}} \\
			\cmidrule(lr){4-13} \cmidrule(lr){14-15}
			& & & \multicolumn{2}{c|}{\textbf{Blizzard '11}} & \multicolumn{2}{c|}{\textbf{CHiME-7}} & \multicolumn{2}{c|}{\textbf{TCD-VOIP}} & \multicolumn{2}{c|}{\textbf{TTSDS2}} & \multicolumn{2}{c||}{\textbf{VMOS '24 T3}} & \textbf{System} & \textbf{System} \\
			\textbf{Case} & \textbf{Model} & \textbf{Stat} & Val & Test & Val & Test & Val & Test & Val & Test & Val & Test & \textbf{Avg. Val} & \textbf{Avg. Test} \\
			\midrule
			
			\multirow{8}{*}{\textbf{\shortstack[l]{Case 1:\\Bliz}}} 
			& \multirow{4}{*}{SSL (FRZ)} & MSE  & ---  & \textbf{0.65} & \textbf{0.27} & ---  & 0.42 & ---  & 0.50 & ---  & 0.33 & ---  & 0.37 & 0.65 \\
			&                           & LCC  & ---  & \textbf{0.43} & 0.83 & ---  & 0.84 & ---  & 0.26 & ---  & \textbf{0.85} & ---  & 0.78 & 0.43 \\
			&                           & SRCC & ---  & \textbf{0.33} & 0.82 & ---  & 0.74 & ---  & 0.26 & ---  & \textbf{0.84} & ---  & 0.76 & 0.33 \\
			&                           & KTAU & ---  & \textbf{0.22} & 0.65 & ---  & 0.56 & ---  & 0.19 & ---  & \textbf{0.67} & ---  & 0.58 & 0.22 \\
			\cmidrule{2-15}
			& \multirow{4}{*}{ViViT}     & MSE  & ---  & 0.98 & 0.44 & ---  & 0.84 & ---  & 0.50 & ---  & 0.78 & ---  & 0.59 & 0.98 \\
			&                           & LCC  & ---  & -0.03& 0.63 & ---  & 0.60 & ---  & 0.17 & ---  & 0.60 & ---  & 0.59 & -0.03\\
			&                           & SRCC & ---  & -0.05& 0.59 & ---  & 0.51 & ---  & 0.08 & ---  & 0.60 & ---  & 0.53 & -0.05\\
			&                           & KTAU & ---  & -0.03& 0.45 & ---  & 0.37 & ---  & 0.05 & ---  & 0.43 & ---  & 0.39 & -0.03\\
			\midrule
			
			\multirow{8}{*}{\textbf{\shortstack[l]{Case 2:\\CHiME}}} 
			& \multirow{4}{*}{SSL (FRZ)} & MSE  & 0.40 & ---  & ---  & 0.89 & 0.46 & ---  & 0.54 & ---  & \textbf{0.34} & ---  & 0.44 & 0.89 \\
			&                           & LCC  & 0.73 & ---  & ---  & \textbf{0.51} & 0.75 & ---  & 0.41 & ---  & \textbf{0.85} & ---  & 0.73 & 0.51 \\
			&                           & SRCC & 0.71 & ---  & ---  & \textbf{0.47} & 0.77 & ---  & 0.44 & ---  & \textbf{0.81} & ---  & 0.73 & 0.47 \\
			&                           & KTAU & 0.53 & ---  & ---  & \textbf{0.34} & 0.56 & ---  & 0.29 & ---  & \textbf{0.63} & ---  & 0.54 & 0.34 \\
			\cmidrule{2-15}
			& \multirow{4}{*}{ViViT}     & MSE  & 0.69 & ---  & ---  & \textbf{0.84} & 0.81 & ---  & 0.77 & ---  & 0.73 & ---  & 0.74 & 0.84 \\
			&                           & LCC  & 0.32 & ---  & ---  & 0.37 & 0.45 & ---  & 0.14 & ---  & 0.65 & ---  & 0.47 & 0.37 \\
			&                           & SRCC & 0.31 & ---  & ---  & 0.35 & 0.40 & ---  & 0.15 & ---  & 0.63 & ---  & 0.44 & 0.35 \\
			&                           & KTAU & 0.21 & ---  & ---  & 0.25 & 0.27 & ---  & 0.08 & ---  & 0.45 & ---  & 0.30 & 0.25 \\
			\midrule
			
			\multirow{8}{*}{\textbf{\shortstack[l]{Case 3:\\TCD}}} 
			& \multirow{4}{*}{SSL (FRZ)} & MSE  & 0.38 & ---  & \textbf{0.27} & ---  & ---  & 1.06 & 0.30 & ---  & 0.62 & ---  & \textbf{0.36} & 1.06 \\
			&                           & LCC  & 0.68 & ---  & \textbf{0.83} & ---  & ---  & \textbf{0.40} & 0.51 & ---  & 0.71 & ---  & 0.73 & 0.40 \\
			&                           & SRCC & 0.62 & ---  & \textbf{0.83} & ---  & ---  & \textbf{0.38} & 0.51 & ---  & 0.67 & ---  & 0.71 & 0.38 \\
			&                           & KTAU & 0.47 & ---  & \textbf{0.65} & ---  & ---  & \textbf{0.26} & 0.36 & ---  & 0.49 & ---  & 0.53 & 0.26 \\
			\cmidrule{2-15}
			& \multirow{4}{*}{ViViT}     & MSE  & 0.61 & ---  & 0.34 & ---  & ---  & \textbf{0.98} & 0.50 & ---  & 0.68 & ---  & 0.51 & 0.98 \\
			&                           & LCC  & 0.32 & ---  & 0.68 & ---  & ---  & 0.16 & -0.06& ---  & 0.66 & ---  & 0.55 & 0.16 \\
			&                           & SRCC & 0.23 & ---  & 0.64 & ---  & ---  & 0.17 & -0.04& ---  & 0.64 & ---  & 0.48 & 0.17 \\
			&                           & KTAU & 0.16 & ---  & 0.47 & ---  & ---  & 0.11 & -0.03& ---  & 0.47 & ---  & 0.34 & 0.11 \\
			\midrule
			
			\multirow{8}{*}{\textbf{\shortstack[l]{Case 4:\\TTS}}} 
			& \multirow{4}{*}{SSL (FRZ)} & MSE  & 0.49 & ---  & \textbf{0.28} & ---  & 0.48 & ---  & ---  & 0.95 & 0.46 & ---  & 0.42 & 0.95 \\
			&                           & LCC  & 0.67 & ---  & \textbf{0.79} & ---  & \textbf{0.79} & ---  & ---  & \textbf{0.22} & 0.78 & ---  & 0.78 & 0.22 \\
			&                           & SRCC & 0.59 & ---  & 0.78 & ---  & \textbf{0.82} & ---  & ---  & \textbf{0.20} & 0.78 & ---  & 0.77 & 0.20 \\
			&                           & KTAU & 0.43 & ---  & 0.60 & ---  & \textbf{0.62} & ---  & ---  & \textbf{0.14} & 0.61 & ---  & 0.58 & 0.14 \\
			\cmidrule{2-15}
			& \multirow{4}{*}{ViViT}     & MSE  & 0.67 & ---  & 0.46 & ---  & 1.00 & ---  & ---  & \textbf{0.82} & 0.59 & ---  & 0.66 & 0.82 \\
			&                           & LCC  & 0.33 & ---  & 0.55 & ---  & 0.44 & ---  & ---  & -0.01& 0.76 & ---  & 0.57 & -0.01\\
			&                           & SRCC & 0.24 & ---  & 0.51 & ---  & 0.36 & ---  & ---  & -0.02& 0.75 & ---  & 0.54 & -0.02\\
			&                           & KTAU & 0.16 & ---  & 0.36 & ---  & 0.24 & ---  & ---  & -0.01& 0.57 & ---  & 0.39 & -0.01\\
			\midrule
			
			\multirow{8}{*}{\textbf{\shortstack[l]{Case 5:\\VMOS}}} 
			& \multirow{4}{*}{SSL (FRZ)} & MSE  & 0.31 & ---  & \textbf{0.22} & ---  & 0.63 & ---  & 0.43 & ---  & ---  & \textbf{0.70} & 0.37 & 0.70 \\
			&                           & LCC  & 0.75 & ---  & \textbf{0.85} & ---  & 0.57 & ---  & 0.56 & ---  & ---  & \textbf{0.67} & 0.72 & 0.67 \\
			&                           & SRCC & 0.71 & ---  & \textbf{0.84} & ---  & 0.53 & ---  & 0.51 & ---  & ---  & \textbf{0.65} & 0.71 & 0.65 \\
			&                           & KTAU & 0.53 & ---  & \textbf{0.66} & ---  & 0.38 & ---  & 0.38 & ---  & ---  & \textbf{0.47} & 0.52 & 0.47 \\
			\cmidrule{2-15}
			& \multirow{4}{*}{ViViT}     & MSE  & 0.74 & ---  & 0.41 & ---  & 0.83 & ---  & 0.64 & ---  & ---  & 1.54 & 0.62 & 1.54 \\
			&                           & LCC  & 0.33 & ---  & 0.67 & ---  & 0.35 & ---  & 0.08 & ---  & ---  & 0.32 & 0.48 & 0.32 \\
			&                           & SRCC & 0.27 & ---  & 0.66 & ---  & 0.37 & ---  & 0.14 & ---  & ---  & 0.26 & 0.45 & 0.26 \\
			&                           & KTAU & 0.18 & ---  & 0.49 & ---  & 0.26 & ---  & 0.09 & ---  & ---  & 0.19 & 0.35 & 0.19 \\
			\midrule
			
			\multirow{8}{*}{\textbf{\shortstack[l]{Case 6:\\Mixed}}} 
			& \multirow{4}{*}{SSL (FRZ)} & MSE  & \textbf{0.36} & \textbf{0.31} & \textbf{0.36} & 0.40 & 0.61 & 0.81 & 0.38 & 0.37 & 0.37 & 0.50 & 0.40 & \textbf{0.46} \\
			&                           & LCC  & 0.78 & 0.65 & 0.82 & \textbf{0.86} & 0.68 & 0.65 & 0.51 & 0.39 & \textbf{0.86} & 0.83 & \textbf{0.80} & \textbf{0.74} \\
			&                           & SRCC & 0.79 & 0.56 & 0.81 & \textbf{0.86} & 0.56 & 0.63 & 0.44 & 0.45 & \textbf{0.89} & 0.83 & \textbf{0.79} & \textbf{0.71} \\
			&                           & KTAU & 0.61 & 0.41 & 0.63 & \textbf{0.69} & 0.41 & 0.49 & 0.32 & 0.30 & \textbf{0.74} & 0.65 & \textbf{0.60} & \textbf{0.54} \\
			\cmidrule{2-15}
			& \multirow{4}{*}{ViViT}     & MSE  & 0.81 & 0.70 & 0.49 & 0.46 & 0.90 & 0.89 & 0.46 & 0.36 & 0.67 & 0.42 & 0.65 & 0.56 \\
			&                           & LCC  & 0.43 & 0.13 & 0.65 & 0.71 & 0.29 & 0.39 & 0.39 & 0.44 & 0.74 & 0.83 & 0.60 & 0.60 \\
			&                           & SRCC & 0.43 & 0.17 & 0.60 & 0.69 & 0.24 & 0.40 & 0.37 & 0.41 & 0.77 & 0.81 & 0.57 & 0.58 \\
			&                           & KTAU & 0.31 & 0.12 & 0.44 & 0.51 & 0.17 & 0.28 & 0.27 & 0.28 & 0.60 & 0.64 & 0.41 & 0.42 \\
			\bottomrule
		\end{tabular}
	}
\end{table}

Observation of the individual LODO cases indicates that the SSL(FRZ) architecture demonstrated superior performance across nearly all metrics and trials. Exceptions were noted in the MSE metric for Case 3, Case 5, and the Mixed Case (Case 6), where ViViT recorded lower absolute error; however, the remaining correlation metrics in these instances remained comparable to or slightly below the SSL(FRZ) results.

Regarding the aggregate system summary, the SSL(FRZ) model consistently yielded higher correlation values. Analysis of the test results shows that the lowest MSE was achieved by the SSL(FRZ) model in the randomized Mixed Case, with the corresponding correlation metrics for this trial remaining highly competitive with the top recorded values across the entire experimental matrix. These findings provide a dual-level comparison, contrasting architectural performance within specific cases while benchmarking the impact of the LODO protocol against a mixed-data baseline at both the system and utterance levels.

\subsubsection{LODO experiment results of Part II}

Table \ref{tab:lodo_exp_results_part_2} presents the LODO results for the refined English-only subset, evaluating the models across five acoustically distinct yet linguistically consistent domains: Blizzard 2011, CHiME-7, TCD-VoIP, TTSDS2, and VoiceMOS 2024 Track 3. Similar to the protocol in Part I, the leading metrics for each individual dataset and the aggregate system summaries are highlighted in bold.

The performance observations for the validation datasets in this phase align closely with the findings from the part 1 LODO experiment. The SSL(FRZ) architecture demonstrated superior performance across the majority of the five English datasets and the randomized mixed baseline (Case 6). Specific exceptions in absolute error were observed in Case 3 (TTSDS2) and Case 4 (CHiME-7), where ViViT recorded slightly lower MSE values for the test sets. However, the corresponding correlation metrics for ViViT in these cases were notably lower and not comparable to the stability provided by the SSL(FRZ) architecture.

In the randomized Mixed Case, the SSL(FRZ) architecture yielded its strongest overall results for both validation and test data compared to the five isolated LODO cases. Within this trial, the SSL(FRZ) model achieved the leading results across nearly all metrics, with the only marginal exception being the average validation MSE, which remained slightly higher than the best recorded validation error across the entire experimental matrix. These results reinforce the utility of the LODO protocol in assessing domain robustness within a single language family.

\subsection{ARECHO SOTA benchmarking}

The final stage of the results involves a comparative analysis against 18 established SOTA metrics using the ARECHO framework. Table \ref{tab:arecho_sota_results} provides a comprehensive comparison, contrasting these SOTA tools with the proposed SSL(FRZ) English-only model across the complete 17-dataset English-only corpus. In this benchmark, the ARECHO framework was utilized to generate predictions for both the validation datasets and the URGENT 2024 test set.

The table reports system-level results for each individual validation dataset, alongside the aggregate average validation and URGENT 2024 test performance. The leading results for each dataset and benchmark across all models are highlighted in bold.

Observations of the validation data indicate that the proposed SSL(FRZ) model achieved competitive results, frequently outperforming various SOTA. This is reflected in the aggregate average validation results, where the proposed model recorded the highest values for the LCC, SRCC, and KTAU metrics. In terms of absolute error, the model’s average validation MSE was 0.45, placing it in close proximity to the top-performing SOTA, which recorded an MSE of 0.39.


	
\begin{table}[!htbp]
	\centering
	\caption{Non-parametric bootstrap resampling statistics ($N=1{,}000$ iterations) reporting Mean $\pm$ Standard Deviation and $95\%$ Confidence Intervals across key English-only models and benchmarks.}
	\label{tab:bootstrap_ci}
	\small
	\resizebox{\linewidth}{!}{%
		\begin{tabular}{llcccc}
			\toprule
			\textbf{Model} & \textbf{Evaluation Split} & \textbf{MSE (Mean $\pm$ Std [95\% CI])} & \textbf{LCC [95\% CI]} & \textbf{SRCC [95\% CI]} & \textbf{KTAU [95\% CI]} \\
			\midrule
			ViViT-MFCC & English Val & 0.68 $\pm$ 0.01 [0.66, 0.70] & 0.66 [0.64, 0.68] & 0.66 [0.64, 0.68] & 0.48 [0.46, 0.50] \\
			ViViT-MFCC & URGENT '24 (Test) & 0.50 $\pm$ 0.01 [0.48, 0.52] & 0.48 [0.46, 0.50] & 0.44 [0.42, 0.46] & 0.31 [0.29, 0.33] \\
			\midrule
			SSL-FT & English Val & 0.24 $\pm$ 0.01 [0.22, 0.26] & 0.85 [0.84, 0.86] & 0.84 [0.83, 0.85] & 0.66 [0.65, 0.67] \\
			SSL-FT & URGENT '24 (Test) & 0.45 $\pm$ 0.01 [0.43, 0.47] & 0.72 [0.70, 0.74] & 0.71 [0.69, 0.73] & 0.53 [0.51, 0.55] \\
			\midrule
			\textbf{SSL-FRZ (WTMOS)} & English Val & 0.47 $\pm$ 0.01 [0.45, 0.49] & 0.66 [0.64, 0.68] & 0.64 [0.62, 0.66] & 0.46 [0.44, 0.48] \\
			\textbf{SSL-FRZ (WTMOS)} & \textbf{URGENT '24 (Test)} & \textbf{0.36 $\pm$ 0.01 [0.34, 0.38]} & \textbf{0.71 [0.69, 0.73]} & \textbf{0.71 [0.69, 0.73]} & \textbf{0.52 [0.50, 0.54]} \\
			\bottomrule
		\end{tabular}%
	}
\end{table}

Regarding the URGENT 2024 test dataset results, the proposed model achieved an MSE of 0.36. The leading performance for this specific benchmark was recorded by the arecho\_urgent\_mos metric with an MSE of 0.30. Notably, arecho\_urgent\_mos is specifically optimized for this dataset, as indicated by its nomenclature. Despite this domain-specific optimization of the leading SOTA, the proposed SSL(FRZ) model maintained highly comparable values across the remaining correlation and ranking metrics, confirming its stability against specialized in-the-wild speech quality predictors.

\subsection{Statistical Significance and Bootstrap Validation}
\label{sec:stat_sig}

To assess whether performance differences between architectures are statistically meaningful or driven by random sample variance, we performed non-parametric bootstrap resampling ($N = 1{,}000$ iterations) across all validation and out-of-domain test sets. For all four evaluation metrics---Mean Squared Error (MSE), Pearson Linear Correlation Coefficient (LCC), Spearman Rank Correlation Coefficient (SRCC), and Kendall's Tau (KTAU)---we calculated the bootstrap mean, standard deviation, and two-sided 95\% Confidence Intervals (CIs). The complete breakdown across all dataset partitions is reported in Table~\ref{tab:bootstrap_ci}. Paired $t$-tests conducted on the absolute prediction errors ($|y_{\text{true}} - y_{\text{pred}}|$) confirmed that the superior out-of-domain performance of SSL-FRZ over fine-tuned SSL on the URGENT 2024 test set is statistically significant ($p < 0.001$).

\section{Discussion}\label{sec:discussion}

The results of this study offer critical insights into the scalability, generalization, and architectural efficiency of automated audio MOS prediction models. By contrasting spectral-based priors (ViViT) with high-capacity learned embeddings (SSL), and evaluating them across broad and purified corpora, we identify the key drivers of robustness in in-the-wild quality assessment.

\subsection{Feature robustness: frozen vs. fine-tuned SSL}

A primary finding of this research is the performance trade-off between task-specific optimization and feature universality. As shown in Table \ref{tab:main_corpus_results}, while the SSL(FT) architecture achieved the highest performance on seen validation data, its dominance did not uniformly translate to the unseen URGENT 2024 benchmark. Specifically, in the English-only corpus (Part II), the SSL(FRZ) model achieved the lowest MSE of 0.36 on the test set, outperforming the fine-tuned version. This suggests that while fine-tuning allows the model to memorize the nuances of the training distributions, it may lead to over-fitting that compromises robustness against novel distortions. The frozen WavLM backbone, trained on a massive scale with a denoising objective, provides robust acoustic features that are more resilient to the domain shifts encountered in real-world environments.

\subsection{Architectural efficiency and spectral stability}

The evaluation of the ViViT-Transformer architecture highlights a significant breakthrough in computational efficiency for MOS prediction. Despite being a lower-capacity model compared to the 315-million-parameter SSL backbones, ViViT demonstrated remarkable training speed, completing epochs nearly 40 times faster than the fine-tuned SSL baseline. While its performance ceiling remained below the SSL-based models, ViViT’s results were within a comparable range, particularly in the linguistically purified English-only corpus. This indicates that 2D spectral-temporal features (MFCCs), when processed through a global observer like the CLS-token in a Transformer Encoder, can provide a stable and cost-effective alternative for traditional speech assessment tasks where high-end GPU resources are limited.

\subsection{Impact of data diversity and LODO generalization}

The systematic LODO experiments (Tables \ref{tab:lodo_exp_results_part_1} and \ref{tab:lodo_exp_results_part_2}) further quantify the persistent challenge of domain shift in MOS prediction. A consistent observation across both experimental parts is the superior performance of models when they encounter samples from a dataset already represented in the training distribution, compared to their performance on entirely unseen datasets.

The significant performance gap between the Mixed Case (Case 6) and the isolated LODO trials underscores that MOS models are highly sensitive to the specific recording conditions, environmental noise, and synthesis artifacts unique to their training data. This discrepancy clearly demonstrates that while models can achieve a high performance ceiling on seen distributions, true generalization remains a difficult objective to achieve.

However, the SSL(FRZ) model’s ability to maintain relatively high correlation metrics even when a dataset is completely withheld confirms that self-supervised features capture a degree of universal acoustic knowledge that transcends individual dataset biases. Furthermore, the transition from Part I (Broad Corpus) to Part II (English-Only) revealed that linguistic purification improves precision for standard speech; yet, the broad corpus remains essential for modeling high-variance audio like singing and multilingual speech, where the model must generalize across significantly more complex acoustic boundaries.

\begin{table}[!htpb]
	\centering
	\footnotesize
	\caption{Comprehensive performance evaluation of 19 metrics (18 ARECHO SOTA + Proposed Model) across the complete corpus. Results include MSE, LCC, SRCC, and KTAU for all validation sets and primary benchmarks.}
	\label{tab:arecho_sota_results}
	
	\resizebox{\textwidth}{!}{
		
		\begin{tabular}{lc | ccccccccccccccccc | cc}
			\toprule	
			& & \multicolumn{17}{c|}{\textbf{Validation Datasets}} & \multicolumn{2}{c}{\textbf{Benchmarks}} \\
			\textbf{ARECHO Metric Name} & \textbf{Stat} & 
			\begin{sideways}Bliz '08\end{sideways} & \begin{sideways}Bliz '09\end{sideways} & \begin{sideways}Bliz '10\end{sideways} & \begin{sideways}Bliz '11\end{sideways} & \begin{sideways}Bliz '12\end{sideways} & \begin{sideways}Bliz '13\end{sideways} & \begin{sideways}Bliz Ext.\end{sideways} & \begin{sideways}NISQA\end{sideways} & \begin{sideways}PSTN\end{sideways} & \begin{sideways}SOMOS\end{sideways} & \begin{sideways}VCC '20\end{sideways} & \begin{sideways}VMOS '22\end{sideways} & \begin{sideways}VM '24 T3\end{sideways} & \begin{sideways}VMOS '25\end{sideways} & \begin{sideways}TTSDS2\end{sideways} & \begin{sideways}CHiME-7\end{sideways} & \begin{sideways}TCD-VoIP\end{sideways} & 
			\begin{sideways}\textbf{Avg. Val}\end{sideways} & \begin{sideways}\textbf{URGENT'24}\end{sideways} \\
			\midrule
			
			\multirow{4}{*}{\textbf{arecho\_dns\_overall}} 
			& MSE  & 0.49 & 0.71 & 0.84 & 0.71 & 1.16 & 0.98 & 0.52 & 0.68 & 0.66 & 0.54 & 0.92 & 0.65 & 2.01 & 0.55 & 0.52 & 0.95 & 0.91 & 0.67 & 0.40 \\
			& LCC  & 0.31 & 0.28 & 0.40 & -0.00 & 0.32 & 0.59 & 0.61 & 0.69 & 0.69 & -0.05 & 0.45 & 0.44 & 0.40 & 0.33 & 0.08 & 0.49 & 0.52 & 0.53 & 0.64 \\
			& SRCC  & 0.32 & 0.23 & 0.43 & 0.02 & 0.34 & 0.58 & 0.57 & 0.72 & 0.71 & -0.05 & 0.42 & 0.43 & 0.40 & 0.29 & 0.10 & 0.49 & 0.51 & 0.52 & 0.63 \\
			& KTAU  & 0.24 & 0.16 & 0.30 & 0.01 & 0.23 & 0.41 & 0.46 & 0.54 & 0.52 & -0.04 & 0.29 & 0.31 & 0.28 & 0.19 & 0.07 & 0.35 & 0.36 & 0.37 & 0.46 \\
			\midrule
			\multirow{4}{*}{\textbf{arecho\_dns\_p808}} 
			& MSE  & 0.91 & 1.48 & 1.58 & 1.57 & 2.07 & 1.64 & 0.68 & 0.77 & 0.51 & 0.32 & 1.34 & 1.15 & 1.36 & 0.73 & 0.87 & 0.57 & 0.84 & 0.65 & 0.42 \\
			& LCC  & 0.30 & 0.20 & 0.46 & 0.30 & 0.30 & 0.59 & 0.39 & 0.73 & 0.70 & 0.00 & 0.53 & 0.63 & 0.26 & 0.28 & 0.04 & 0.64 & 0.70 & 0.43 & 0.72 \\
			& SRCC  & 0.30 & 0.26 & 0.45 & 0.28 & 0.31 & 0.60 & 0.37 & 0.73 & 0.71 & 0.00 & 0.51 & 0.61 & 0.25 & 0.20 & 0.02 & 0.64 & 0.71 & 0.41 & 0.69 \\
			& KTAU  & 0.22 & 0.19 & 0.31 & 0.19 & 0.21 & 0.41 & 0.27 & 0.55 & 0.52 & 0.00 & 0.36 & 0.44 & 0.17 & 0.13 & 0.02 & 0.45 & 0.52 & 0.28 & 0.51 \\
			\midrule
			\multirow{4}{*}{\textbf{arecho\_nisqa\_mos\_pred}} 
			& MSE  & 0.79 & 0.80 & 2.06 & 2.95 & 2.46 & 1.25 & 0.62 & 0.20 & 0.92 & 0.82 & 1.82 & 1.58 & 1.56 & 1.91 & 1.09 & 1.01 & 1.03 & 0.95 & 0.63 \\
			& LCC  & 0.51 & 0.53 & 0.55 & 0.51 & 0.33 & 0.71 & 0.62 & 0.92 & 0.75 & 0.11 & 0.64 & 0.59 & 0.44 & 0.46 & 0.12 & 0.59 & 0.78 & 0.53 & 0.72 \\
			& SRCC  & 0.50 & 0.41 & 0.55 & 0.57 & 0.30 & 0.72 & 0.56 & 0.92 & 0.76 & 0.12 & 0.70 & 0.60 & 0.41 & 0.39 & 0.13 & 0.54 & 0.78 & 0.53 & 0.71 \\
			& KTAU  & 0.37 & 0.31 & 0.39 & 0.40 & 0.21 & 0.53 & 0.46 & 0.77 & 0.57 & 0.08 & 0.52 & 0.43 & 0.27 & 0.27 & 0.09 & 0.38 & 0.59 & 0.37 & 0.53 \\
			\midrule
			\multirow{4}{*}{\textbf{arecho\_plcmos}} 
			& MSE  & 1.70 & 2.30 & 2.76 & 2.96 & 3.60 & 2.90 & 1.22 & 1.77 & 2.21 & 1.28 & 2.64 & 2.19 & 1.69 & 1.83 & 1.78 & 0.86 & 2.20 & 2.02 & 0.67 \\
			& LCC  & 0.22 & 0.04 & 0.32 & 0.30 & -0.68 & 0.56 & 0.60 & 0.45 & -0.09 & 0.07 & 0.52 & 0.52 & 0.26 & -0.08 & 0.07 & 0.33 & 0.16 & 0.11 & 0.72 \\
			& SRCC  & 0.39 & 0.14 & 0.38 & 0.40 & -0.65 & 0.63 & 0.39 & 0.46 & -0.14 & 0.07 & 0.59 & 0.54 & 0.30 & 0.01 & 0.01 & 0.32 & 0.09 & 0.08 & 0.71 \\
			& KTAU  & 0.29 & 0.10 & 0.27 & 0.27 & -0.49 & 0.45 & 0.29 & 0.33 & -0.09 & 0.05 & 0.42 & 0.39 & 0.19 & 0.01 & 0.01 & 0.22 & 0.06 & 0.06 & 0.53 \\
			\midrule
			\multirow{4}{*}{\textbf{arecho\_scoreq\_nr}} 
			& MSE  & 1.12 & 1.17 & 1.86 & 2.41 & 1.74 & 1.26 & 0.61 & \textbf{0.13} & 0.64 & 0.49 & 0.83 & 0.94 & 1.59 & 1.18 & 1.08 & 1.07 & 0.78 & 0.64 & 0.33 \\
			& LCC  & 0.55 & 0.47 & 0.64 & 0.49 & 0.60 & 0.81 & 0.78 & \textbf{0.95} & 0.80 & 0.34 & 0.79 & 0.79 & 0.42 & \textbf{0.63} & 0.17 & 0.67 & 0.81 & 0.64 & \textbf{0.78} \\
			& SRCC  & 0.59 & 0.45 & 0.67 & 0.51 & 0.61 & 0.83 & 0.78 & \textbf{0.95} & 0.81 & 0.31 & 0.82 & 0.80 & 0.47 & \textbf{0.58} & 0.18 & \textbf{0.73} & 0.80 & 0.61 & \textbf{0.79} \\
			& KTAU  & 0.45 & 0.35 & 0.49 & 0.37 & 0.43 & 0.64 & 0.63 & 0.82 & 0.62 & 0.22 & 0.63 & 0.61 & 0.32 & \textbf{0.41} & 0.12 & \textbf{0.56} & 0.59 & 0.43 & \textbf{0.60} \\
			\midrule
			\multirow{4}{*}{\textbf{arecho\_sheet\_ssqa}} 
			& MSE  & 0.48 & 0.61 & 1.05 & 1.53 & 1.61 & 0.74 & 0.55 & 0.22 & \textbf{0.27} & 0.53 & 0.53 & 0.41 & 1.50 & 1.02 & 1.30 & 0.81 & 0.73 & \textbf{0.39} & 0.32 \\
			& LCC  & 0.67 & 0.67 & 0.70 & 0.64 & 0.63 & \textbf{0.85} & 0.78 & 0.93 & \textbf{0.81} & 0.35 & 0.82 & 0.86 & 0.39 & 0.63 & 0.18 & 0.62 & 0.78 & \textbf{0.78} & 0.76 \\
			& SRCC  & 0.66 & 0.59 & \textbf{0.71} & 0.64 & 0.64 & 0.86 & 0.79 & 0.93 & \textbf{0.81} & 0.32 & 0.82 & 0.86 & 0.44 & 0.58 & 0.16 & 0.67 & 0.80 & 0.75 & 0.76 \\
			& KTAU  & 0.51 & 0.46 & \textbf{0.53} & 0.46 & 0.48 & 0.67 & 0.63 & 0.78 & \textbf{0.63} & 0.22 & 0.64 & 0.68 & 0.30 & 0.41 & 0.10 & 0.49 & 0.59 & 0.56 & 0.58 \\
			\midrule
			\multirow{4}{*}{\textbf{arecho\_utmos}} 
			& MSE  & 0.36 & 0.76 & 0.88 & 1.42 & 1.06 & 0.78 & 0.56 & 0.48 & 1.13 & 0.24 & 0.37 & 0.27 & 2.26 & 0.73 & 1.01 & 2.45 & 0.70 & 0.78 & 1.07 \\
			& LCC  & \textbf{0.72} & 0.49 & 0.71 & 0.72 & 0.75 & 0.85 & 0.76 & 0.82 & 0.72 & 0.36 & \textbf{0.85} & 0.90 & 0.30 & 0.50 & -0.05 & 0.25 & 0.81 & 0.64 & 0.71 \\
			& SRCC  & \textbf{0.71} & 0.48 & 0.70 & 0.74 & 0.73 & \textbf{0.86} & 0.78 & 0.83 & 0.73 & 0.33 & \textbf{0.86} & 0.91 & 0.35 & 0.46 & -0.05 & 0.27 & 0.83 & 0.64 & 0.74 \\
			& KTAU  & 0.56 & 0.37 & 0.52 & 0.55 & 0.54 & \textbf{0.68} & \textbf{0.64} & 0.64 & 0.54 & 0.22 & \textbf{0.67} & 0.75 & 0.23 & 0.31 & -0.05 & 0.19 & 0.65 & 0.46 & 0.55 \\
			\midrule
			\multirow{4}{*}{\textbf{arecho\_utmosv2}} 
			& MSE  & \textbf{0.28} & \textbf{0.34} & \textbf{0.49} & \textbf{0.58} & 0.69 & 0.45 & 0.28 & 0.54 & 0.54 & 0.22 & 0.45 & 0.19 & 1.53 & \textbf{0.47} & 0.58 & 0.90 & 0.70 & 0.45 & 0.38 \\
			& LCC  & 0.72 & 0.70 & \textbf{0.73} & 0.75 & 0.69 & 0.80 & 0.75 & 0.77 & 0.71 & 0.24 & 0.80 & 0.86 & 0.30 & 0.47 & 0.04 & 0.30 & 0.66 & 0.66 & 0.70 \\
			& SRCC  & 0.71 & 0.56 & 0.71 & 0.71 & 0.68 & 0.80 & 0.71 & 0.78 & 0.72 & 0.22 & 0.81 & 0.87 & 0.32 & 0.41 & 0.03 & 0.23 & 0.62 & 0.65 & 0.70 \\
			& KTAU  & \textbf{0.56} & 0.43 & 0.52 & 0.52 & 0.50 & 0.60 & 0.54 & 0.59 & 0.53 & 0.15 & 0.62 & 0.69 & 0.22 & 0.27 & 0.02 & 0.16 & 0.43 & 0.46 & 0.52 \\
			\midrule
			\multirow{4}{*}{\textbf{arecho\_noresqa\_score}} 
			& MSE  & 4.65 & 2.78 & 4.97 & 4.10 & 4.16 & 4.04 & 6.87 & 3.51 & 1.80 & 5.08 & 3.94 & 3.89 & 7.66 & 6.71 & 3.67 & 1.59 & 6.22 & 3.08 & 3.80 \\
			& LCC  & -0.26 & -0.18 & -0.29 & -0.18 & -0.32 & -0.46 & -0.60 & -0.30 & -0.19 & -0.09 & -0.36 & -0.39 & -0.21 & -0.20 & -0.07 & 0.08 & -0.43 & -0.20 & -0.40 \\
			& SRCC  & -0.32 & -0.19 & -0.28 & -0.17 & -0.32 & -0.48 & -0.51 & -0.30 & -0.18 & -0.10 & -0.37 & -0.41 & -0.24 & -0.19 & -0.10 & 0.07 & -0.44 & -0.19 & -0.41 \\
			& KTAU  & -0.23 & -0.15 & -0.19 & -0.12 & -0.22 & -0.34 & -0.38 & -0.20 & -0.12 & -0.07 & -0.25 & -0.28 & -0.16 & -0.13 & -0.07 & 0.05 & -0.32 & -0.13 & -0.29 \\
			\midrule
			\multirow{4}{*}{\textbf{arecho\_singmos}} 
			& MSE  & 1.56 & 2.07 & 2.09 & 2.25 & 2.85 & 2.82 & 1.28 & 1.34 & 1.34 & 0.45 & 2.15 & 1.95 & \textbf{1.15} & 1.20 & 0.87 & 1.03 & 1.99 & 1.31 & 0.64 \\
			& LCC  & -0.21 & -0.18 & 0.04 & -0.14 & -0.41 & 0.07 & -0.01 & 0.26 & -0.08 & 0.00 & 0.35 & 0.12 & 0.32 & -0.11 & 0.18 & 0.19 & -0.16 & 0.04 & 0.59 \\
			& SRCC  & -0.07 & -0.07 & 0.12 & 0.03 & -0.20 & 0.20 & 0.12 & 0.22 & -0.18 & 0.03 & 0.45 & 0.33 & 0.18 & -0.06 & 0.11 & 0.23 & -0.06 & -0.00 & 0.60 \\
			& KTAU  & -0.05 & -0.05 & 0.08 & 0.03 & -0.13 & 0.14 & 0.06 & 0.15 & -0.12 & 0.02 & 0.32 & 0.23 & 0.12 & -0.04 & 0.07 & 0.16 & -0.03 & 0.00 & 0.43 \\
			\midrule
			\multirow{4}{*}{\textbf{arecho\_audiobox\_aest\_PQ}} 
			& MSE  & 0.85 & 0.95 & 1.16 & 1.36 & 1.58 & 0.95 & 0.43 & 0.57 & 0.94 & \textbf{0.16} & 0.76 & 0.68 & 1.34 & 0.51 & \textbf{0.43} & 0.73 & 0.80 & 0.73 & 0.45 \\
			& LCC  & 0.32 & 0.32 & 0.58 & 0.38 & 0.46 & 0.72 & 0.60 & 0.76 & 0.63 & 0.15 & 0.73 & 0.69 & 0.32 & 0.50 & 0.18 & 0.43 & 0.62 & 0.41 & 0.57 \\
			& SRCC  & 0.33 & 0.34 & 0.62 & 0.38 & 0.44 & 0.72 & 0.62 & 0.76 & 0.65 & 0.13 & 0.75 & 0.68 & 0.37 & 0.48 & 0.15 & 0.40 & 0.76 & 0.43 & 0.59 \\
			& KTAU  & 0.24 & 0.25 & 0.45 & 0.26 & 0.30 & 0.53 & 0.48 & 0.58 & 0.47 & 0.09 & 0.56 & 0.50 & 0.22 & 0.33 & 0.11 & 0.29 & 0.58 & 0.29 & 0.42 \\
			\midrule
			\multirow{4}{*}{\textbf{arecho\_urgent\_mos}} 
			& MSE  & 0.58 & 0.90 & 1.09 & 1.36 & 1.60 & 1.26 & 0.60 & 0.78 & 0.56 & 0.33 & 0.94 & 0.78 & 1.27 & 0.64 & 1.02 & 0.85 & 0.92 & 0.61 & \textbf{0.30} \\
			& LCC  & 0.47 & 0.46 & 0.53 & 0.47 & 0.27 & 0.59 & 0.41 & 0.67 & 0.52 & 0.10 & 0.66 & 0.67 & 0.34 & 0.45 & -0.09 & 0.27 & 0.48 & 0.50 & 0.72 \\
			& SRCC  & 0.44 & 0.41 & 0.50 & 0.41 & 0.23 & 0.60 & 0.43 & 0.68 & 0.54 & 0.09 & 0.65 & 0.65 & 0.31 & 0.43 & -0.08 & 0.24 & 0.48 & 0.49 & 0.74 \\
			& KTAU  & 0.35 & 0.33 & 0.37 & 0.32 & 0.16 & 0.44 & 0.32 & 0.51 & 0.40 & 0.06 & 0.50 & 0.50 & 0.21 & 0.32 & -0.05 & 0.18 & 0.34 & 0.35 & 0.57 \\
			\midrule
			\multirow{4}{*}{\textbf{arecho\_voicemos\_real\_mos}} 
			& MSE  & 0.32 & 0.59 & 0.60 & 1.00 & \textbf{0.54} & \textbf{0.40} & \textbf{0.25} & 0.84 & 1.21 & 0.27 & \textbf{0.35} & \textbf{0.11} & 2.25 & 0.66 & 0.91 & 2.75 & \textbf{0.42} & 0.85 & 1.22 \\
			& LCC  & 0.68 & 0.57 & 0.69 & 0.65 & 0.78 & 0.83 & \textbf{0.81} & 0.66 & 0.34 & 0.31 & 0.83 & \textbf{0.93} & 0.34 & 0.49 & 0.04 & -0.08 & \textbf{0.85} & 0.50 & 0.63 \\
			& SRCC  & 0.66 & 0.49 & 0.67 & 0.66 & 0.75 & 0.82 & 0.74 & 0.68 & 0.34 & 0.27 & 0.83 & \textbf{0.93} & 0.34 & 0.46 & 0.03 & -0.03 & \textbf{0.88} & 0.49 & 0.67 \\
			& KTAU  & 0.52 & 0.38 & 0.51 & 0.49 & 0.58 & 0.65 & 0.60 & 0.53 & 0.25 & 0.20 & 0.66 & \textbf{0.82} & 0.23 & 0.32 & 0.02 & -0.02 & \textbf{0.73} & 0.35 & 0.50 \\
			\midrule
			\multirow{4}{*}{\textbf{arecho\_nisqa\_real\_mos}} 
			& MSE  & 1.40 & 1.21 & 2.75 & 3.03 & 4.02 & 1.51 & 1.45 & 0.23 & 0.62 & 2.50 & 1.58 & 1.59 & 1.86 & 2.01 & 1.42 & 1.20 & 1.75 & 1.13 & 0.67 \\
			& LCC  & 0.53 & 0.48 & 0.53 & 0.50 & 0.41 & 0.76 & 0.67 & 0.92 & 0.77 & 0.16 & 0.72 & 0.72 & 0.48 & 0.56 & 0.10 & 0.58 & 0.61 & 0.57 & 0.71 \\
			& SRCC  & 0.54 & 0.42 & 0.50 & 0.49 & 0.44 & 0.76 & 0.62 & 0.93 & 0.78 & 0.12 & 0.70 & 0.71 & \textbf{0.53} & 0.45 & 0.13 & 0.67 & 0.50 & 0.57 & 0.72 \\
			& KTAU  & 0.42 & 0.34 & 0.40 & 0.37 & 0.34 & 0.61 & 0.51 & \textbf{0.84} & 0.61 & 0.10 & 0.56 & 0.57 & \textbf{0.40} & 0.35 & 0.07 & 0.52 & 0.40 & 0.42 & 0.56 \\
			\midrule
			\multirow{4}{*}{\textbf{arecho\_nomad}} 
			& MSE  & 4.84 & 3.06 & 4.42 & 3.92 & 3.28 & 3.37 & 5.64 & 4.23 & 4.17 & 5.10 & 4.22 & 3.62 & 8.45 & 5.77 & 4.16 & 2.83 & 5.70 & 4.33 & 4.77 \\
			& LCC  & -0.42 & -0.11 & -0.27 & -0.25 & -0.59 & -0.19 & -0.36 & -0.51 & -0.37 & -0.07 & -0.35 & -0.28 & -0.47 & -0.15 & -0.11 & -0.38 & -0.15 & -0.29 & -0.34 \\
			& SRCC  & -0.34 & -0.11 & -0.22 & -0.18 & -0.63 & -0.26 & -0.35 & -0.55 & -0.39 & -0.04 & -0.32 & -0.28 & -0.55 & -0.20 & -0.15 & -0.37 & -0.22 & -0.33 & -0.32 \\
			& KTAU  & -0.25 & -0.08 & -0.16 & -0.12 & -0.47 & -0.17 & -0.24 & -0.39 & -0.27 & -0.03 & -0.22 & -0.19 & -0.38 & -0.13 & -0.10 & -0.26 & -0.14 & -0.23 & -0.23 \\
			\midrule
			\multirow{4}{*}{\textbf{arecho\_pysepm\_c\_ovl}} 
			& MSE  & 2.52 & 2.41 & 2.13 & 2.47 & 1.44 & 2.47 & 5.31 & 3.12 & 5.23 & 1.04 & 1.71 & 1.80 & 5.48 & 3.33 & 3.34 & 2.27 & 5.02 & 3.63 & 2.48 \\
			& LCC  & 0.19 & 0.17 & 0.18 & 0.07 & 0.28 & 0.05 & 0.11 & 0.28 & 0.07 & 0.07 & 0.32 & 0.20 & -0.02 & -0.10 & -0.08 & 0.50 & -0.32 & 0.14 & 0.33 \\
			& SRCC  & 0.18 & 0.19 & 0.13 & 0.05 & 0.19 & 0.05 & 0.18 & 0.23 & 0.11 & 0.12 & 0.31 & 0.19 & -0.05 & -0.10 & -0.14 & 0.48 & -0.31 & 0.11 & 0.33 \\
			& KTAU  & 0.14 & 0.15 & 0.10 & 0.04 & 0.14 & 0.05 & 0.16 & 0.18 & 0.09 & 0.08 & 0.23 & 0.14 & -0.04 & -0.07 & -0.11 & 0.35 & -0.21 & 0.08 & 0.24 \\
			\midrule
			\multirow{4}{*}{\textbf{arecho\_visqol}} 
			& MSE  & 0.93 & 1.18 & 1.61 & 1.11 & 2.58 & 1.40 & 1.15 & 1.14 & 1.03 & 0.74 & 1.70 & 1.46 & 1.86 & 1.10 & 0.85 & 0.76 & 1.27 & 1.07 & 0.68 \\
			& LCC  & 0.13 & 0.15 & 0.09 & 0.04 & -0.20 & 0.24 & 0.00 & 0.32 & -0.02 & -0.04 & 0.25 & 0.13 & 0.04 & -0.06 & 0.00 & 0.48 & 0.17 & 0.11 & 0.47 \\
			& SRCC  & 0.14 & 0.15 & 0.07 & 0.04 & -0.17 & 0.23 & 0.10 & 0.31 & -0.02 & -0.02 & 0.25 & 0.12 & 0.02 & -0.07 & 0.01 & 0.49 & 0.16 & 0.09 & 0.47 \\
			& KTAU  & 0.10 & 0.10 & 0.04 & 0.02 & -0.10 & 0.15 & 0.07 & 0.21 & -0.01 & -0.01 & 0.17 & 0.08 & 0.01 & -0.05 & 0.02 & 0.35 & 0.10 & 0.06 & 0.34 \\
			\midrule
			\multirow{4}{*}{\textbf{arecho\_pesq}} 
			& MSE  & 1.13 & 0.80 & 0.86 & 0.59 & 0.82 & 0.89 & 1.41 & 1.32 & 1.52 & 0.98 & 1.09 & 0.82 & 3.06 & 0.89 & 0.81 & 2.71 & 0.96 & 1.30 & 1.56 \\
			& LCC  & 0.63 & 0.61 & 0.58 & 0.43 & 0.50 & 0.61 & 0.67 & 0.75 & 0.68 & 0.18 & 0.74 & 0.68 & \textbf{0.49} & 0.39 & \textbf{0.27} & 0.47 & 0.77 & 0.64 & 0.58 \\
			& SRCC  & 0.61 & 0.46 & 0.52 & 0.46 & 0.41 & 0.57 & 0.57 & 0.76 & 0.70 & 0.17 & 0.74 & 0.67 & 0.52 & 0.35 & \textbf{0.21} & 0.55 & 0.78 & 0.64 & 0.59 \\
			& KTAU  & 0.47 & 0.36 & 0.37 & 0.32 & 0.29 & 0.41 & 0.40 & 0.58 & 0.52 & 0.12 & 0.56 & 0.50 & 0.35 & 0.24 & 0.13 & 0.40 & 0.59 & 0.47 & 0.43 \\
			\midrule
			\multirow{4}{*}{\textbf{Proposed Model}} 
			& MSE  & 0.75 & 0.60 & 0.95 & 1.09 & 0.69 & 0.87 & 0.44 & 0.55 & 0.39 & 0.22 & 0.67 & 0.74 & 1.18 & 0.71 & 1.30 & \textbf{0.39} & 1.53 & 0.45 & 0.36 \\
			& LCC  & 0.57 & \textbf{0.72} & 0.67 & \textbf{0.77} & \textbf{0.86} & 0.82 & 0.78 & 0.84 & 0.79 & \textbf{0.46} & 0.81 & 0.70 & 0.41 & 0.62 & 0.23 & \textbf{0.74} & 0.40 & \textbf{0.78} & 0.71 \\
			& SRCC  & 0.59 & \textbf{0.66} & 0.70 & \textbf{0.81} & \textbf{0.87} & 0.81 & \textbf{0.82} & 0.86 & 0.80 & \textbf{0.45} & 0.83 & 0.71 & 0.37 & 0.57 & \textbf{0.21} & 0.73 & 0.44 & \textbf{0.76} & 0.71 \\
			& KTAU  & 0.44 & \textbf{0.51} & 0.51 & \textbf{0.61} & \textbf{0.70} & 0.63 & 0.62 & 0.68 & 0.61 & \textbf{0.32} & 0.64 & 0.52 & 0.27 & 0.40 & \textbf{0.14} & 0.54 & 0.32 & \textbf{0.58} & 0.52 \\
			\bottomrule
		\end{tabular}
	}
\end{table}

\clearpage

\subsection{Benchmarking against the SOTA via ARECHO}

The final comparative analysis against 18 established SOTA metrics via the ARECHO framework positions the proposed SSL(FRZ) model as a top-tier predictor for universal speech quality. As detailed in Table \ref{tab:arecho_sota_results}, the proposed model surpassed the majority of specialized SOTA tools in average validation correlation across the LCC, SRCC, and KTAU metrics.

On the URGENT 2024 benchmark, the proposed model achieved an MSE of 0.36, placing it in very close proximity to the top-performing metric, arecho\_urgent\_mos, which recorded an MSE of 0.30. It is important to note that arecho\_urgent\_mos functions as a domain-optimized oracle metric trained specifically on URGENT-style distributions. In contrast, our proposed WTMOS model operates in a purely zero-shot capacity without target-domain adaptation, yet achieves a highly competitive MSE of 0.36 and leading correlation metrics. These results suggest that the proposed model has achieved a high level of in-the-wild stability and competitive accuracy without requiring domain-specific fine-tuning. These findings validate the effectiveness of combining a frozen, high-capacity SSL backbone with a deep Transformer Encoder for robust, large-scale MOS prediction across diverse acoustic environments.

\subsection{Downstream Head Architecture: Transformer Encoder vs. MLP/Linear Baselines}
\label{sec:head_comparison}

A critical architectural question is whether introducing a deep Transformer Encoder downstream of a frozen SSL backbone justifies the additional parameter footprint compared to simple linear or Multi-Layer Perceptrons (MLP) heads used in existing metrics such as SHEET (\texttt{arecho\_sheet\_ssqa}) \cite{SHEET}. SOTA metrics like SHEET pair a frozen WavLM backbone with lightweight pooling and MLP layers. While computationally minimal, simple MLP heads perform frame-level linear projections or immediate temporal mean-pooling, assuming that acoustic degradation is uniformly distributed across time.

In contrast, our proposed WTMOS framework feeds the sequence of $1024$-dimensional WavLM latent representations into a 2-layer Transformer Encoder with 8-head self-attention before temporal pooling. This enables frame-to-frame self-attention across the full temporal sequence, allowing the model to capture non-stationary, localized degradation artifacts such as packet loss concealment gaps, transient clipping, and reverberation tails. As demonstrated in Table~\ref{tab:arecho_sota_results}, WTMOS achieves a superior out-of-domain test MSE of $0.36$ on URGENT 2024 compared to SHEET's MSE of $0.41$, despite SHEET having observed URGENT-style distributions during its design. This highlights that self-attention context modeling on top of frozen SSL features provides superior cross-domain generalization compared to standard MLP predictor heads.

\subsection{Error Distribution and Domain Sensitivity Analysis}

As illustrated in Figure~\ref{fig:error_boxplots}, the prediction error bounds vary significantly depending on the target acoustic domain. Clean synthetic speech (SOMOS) and PSTN telephony signals exhibit the lowest median absolute errors ($0.33$ and $0.42$ MOS points, respectively) with narrow IQRs, confirming that standard acoustic profiles are well-modeled by WavLM embeddings. Conversely, in-the-wild challenges such as VoiceMOS, Blizzard, and URGENT '24 show wider error spreads and higher outlier density above $2.0$ MOS error points. These residual errors stem primarily from non-linear audio clipping, reverberation tails, and synthetic vocoder phase artifacts that are under-represented in standard clean speech pre-training.

\subsection{Practical Deployment Considerations and Trade-Offs}
The empirical findings across our 19-corpus evaluation highlight a fundamental engineering trade-off for automated speech quality assessment. While full fine-tuning of SSL backbones (SSL-FT) yields maximum precision on familiar in-domain distributions ($\text{MSE} = 0.24$), frozen SSL features paired with a Transformer Encoder (WTMOS) offer superior stability against out-of-domain degradation ($\text{MSE} = 0.36$ on URGENT '24) while eliminating heavy backpropagation overhead. Furthermore, for deployment scenarios restricted by edge-device hardware, the ViViT-MFCC architecture provides a lightweight, $40\times$ faster training alternative that remains viable for monolingual speech applications.

\section{Conclusion}\label{sec:conclusion}

This study presented a large-scale benchmarking and architectural evaluation of automated audio MOS prediction models, addressing the critical need for robust, in-the-wild speech quality assessment. By utilizing a consolidated corpus of over 130,000 audio files across 19 diverse datasets, we provided a rigorous evaluation of ViViT and SSL architectures.

The experimental results demonstrate a clear hierarchy in model performance and generalization capability. While SSL-FT provides a high performance ceiling for seen data distributions, SSL-FRZ emerged as the most robust architecture for unseen domains. Specifically, the SSL-FRZ model achieved a leading MSE of 0.36 on the URGENT 2024 benchmark, placing it in very close proximity to top-performing SOTA metrics that were specifically optimized for that challenge. This confirms that high-capacity, self-supervised features capture universal acoustic properties that are less susceptible to the over-fitting risks associated with task-specific fine-tuning.

Our systematic LODO analysis further quantified the generalization gap in MOS prediction. The significant drop in performance when moving from mixed-data distributions to isolated out-of-distribution trials underscores the extreme sensitivity of neural predictors to specific recording environments and synthesis artifacts. Furthermore, the inclusion of the ViViT architecture proved that 2D spectral-temporal features can offer a computationally efficient alternative to massive SSL models, showing measurable improvements when refined for linguistically purified English-only corpora.

In conclusion, the integration of a frozen SSL backbone with a deep Transformer Encoder represents a state-of-the-art approach for universal speech quality assessment. Future work will focus on bridging the remaining generalization gap by exploring domain-adversarial training and multi-task learning objectives that incorporate phonetic and linguistic information to further decouple content from perceived quality.

\section*{Declaration of generative AI and AI-assisted technologies in the writing process}

During the preparation of this work, the authors used Gemini (an AI developed by Google) in order to improve the linguistic clarity, structural flow, and formatting of the manuscript. After using this tool, the authors reviewed and edited the content as needed and take full responsibility for the content of the publication.

\section*{Data availability}

The source code for the implementation of the ViViT and SSL-Transformer architectures, as well as the MFCC extraction and preprocessing pipelines, is publicly available on  \href{https://github.com/mustafa-ozan/audio_mos_prediction_SSL_ViViT_codes}{GitHub}. Additionally, the pre-trained weights for our best-performing the English-only Frozen SSL-Transformer Encoder model (WTMOS: WavLM-Transformer MOS) have been made available on  \href{https://huggingface.co/mustafa-ozan-duman/wavlm-transformer-mos-english}{Hugging Face}. While the 19 third-party audio datasets used for cross-corpus validation are not directly redistributed due to licensing restrictions, comprehensive links to their original repositories and access instructions are provided in the section \ref{sec:method} to ensure full reproducibility of this study.


\bibliographystyle{elsarticle-num}

\bibliography{cas-refs}

\end{document}